\documentclass[aps,pra,twocolumn,showpacs,letterpaper,superscriptaddress]{revtex4-1}

\usepackage{graphicx,dcolumn,longtable,epsfig}
\usepackage[usenames]{color}
\usepackage{amssymb}
\usepackage{amsmath}
\usepackage{epstopdf}
\usepackage{bm}
\usepackage{footnote}
\usepackage{float}
\usepackage{subfigure}
\usepackage{color}

\input epsf.tex

\def\jpb#1#2#3{J.~Phys.~B:~{\bf #1},\ #2\ (#3)}

\def\pra#1#2#3{Phys.~Rev.~A~{\bf #1},\ #2\ (#3)}
\def\prb#1#2#3{Phys.~Rev.~B~{\bf #1},\ #2\ (#3)}

\def\rmp#1#2#3{Rev.~Mod.~Phys.~{\bf #1},\ #2\ (#3)}
\def\prl#1#2#3{Phys.~Rev.~Lett.~{\bf #1},\ #2\ (#3)}
\def\sci#1#2#3{Science~{\bf #1},\ #2\ (#3)}

\def\njp#1#2#3{New~J.~Phys.~{\bf #1},\ #2\ (#3)}

\def\rmp#1#2#3{Rev.~Mod.~Phys.~{\bf #1},\ #2\ (#3)}

\def\etal{{\it et al.}}
\def\bea{\begin{eqnarray}}
\def\eea{\end{eqnarray}}
\def\be{\begin{equation}}
\def\ee{\end{equation}}

\begin{document}
\setlength{\parskip}{0pt}
\title{Quantum phases of hard-core dipolar bosons in coupled 1D optical lattices}
\author{ A.~Safavi-Naini}
\affiliation{Department of Physics, Massachusetts Institute of Technology, Cambridge, Massachusetts, 02139, USA}
\affiliation{ITAMP, Harvard-Smithsonian Center for Astrophysics, Cambridge, Massachusetts, 02138, USA}

\author{ B.~Capogrosso-Sansone }
\affiliation{Homer L. Dodge Department of Physics and Astronomy, The University of Oklahoma, Norman, Oklahoma 73019, USA }
\author{A. Kuklov}
\affiliation{CUNY}

\begin{abstract}
Hard-core dipolar bosons trapped in a parallel stack of $N\geq 2$ 1D optical lattices (tubes) can develop several phases made of composites of particles from different tubes:  superfluids, supercounterfluids and insulators as well as mixtures of those. Bosonization analysis shows that these phases are threshold-less with respect to the dipolar interaction, with the key ``control knob" being filling factors in each tube, provided the inter-tube tunneling is suppressed. The effective {\it ab-initio} quantum Monte Carlo algorithm capturing these phases is introduced and some results are presented. 
\end{abstract}

\maketitle
\section{Introduction}

The unprecedented level of control in ultra-cold atom experiments has allowed for the realization of paradigmatic condensed matter models \cite{Zoller, Bloch_RMP}. In these systems the inter-particle interactions can be tuned by varying the scattering length through Feshbach resonances and the atoms can be trapped in various geometries~\cite{Sengstock}. Such a flexibility makes ultra-cold atoms an almost ideal candidate for the study of strongly correlated many-body quantum systems as well as a playground for emerging new states of matter.  These, in particular, include paired superfluids. Recent experimental success in trapping ultra-cold bosonic atomic mixtures \cite{Bloch_RMP,mixture1,mixture2, mixture3} has rendered the study of pairing between components very timely. 
The impressively rapid experimental progress towards controlling polar molecules~\cite{pol_mol_exp1, pol_mol_exp2, pol_mol_exp3, pol_mol_exp4, pol_mol_exp5, pol_mol_exp6, pol_mol_exp7, pol_mol_exp8,lattice_spin_pol_mol_exp1}
gives hope for accessing quantum many body systems with long range and anisotropic interaction in a very near future~\cite{lattice_spin_pol_mol_exp1}.

A prominent example of bosonic systems currently available experimentally consists of an array of coupled one-dimensional tubes, with the interaction between tubes provided by dipolar forces. In the absence of the inter-tube tunneling, this system can be relevant to multi-component atomic mixtures. In general, when such a tunneling is finite, it represents coupled spin chains, which is one of the central topics in low-dimensional condensed matter physics~\cite{Giamarchi}. 

\begin{figure}[h]\label{fig:setup}
\begin{center}
\includegraphics[width=0.35\textwidth]{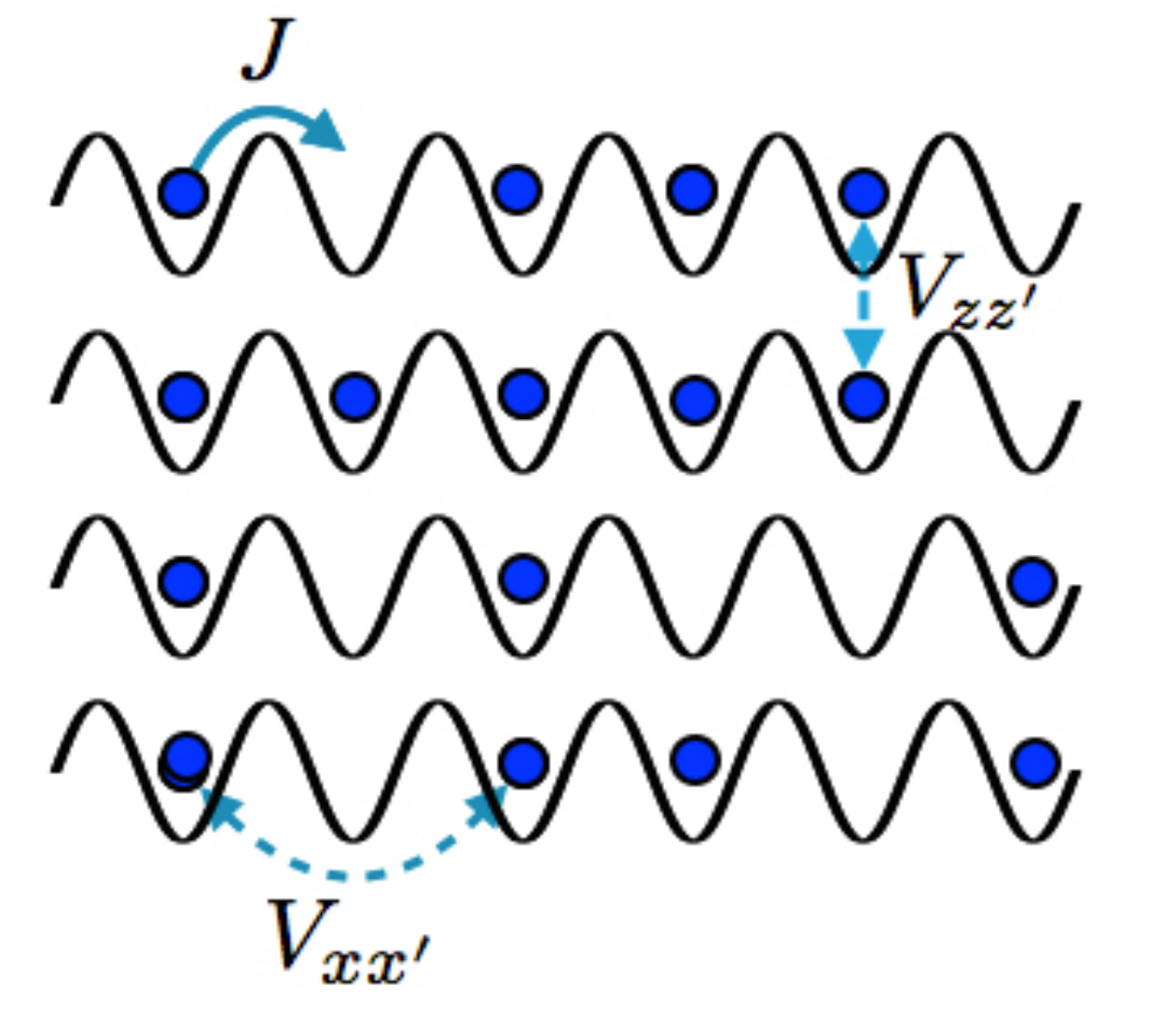}
\caption{A sketch of the system studied: hard core bosons (solid circles) can tunnel (as marked by solid arrow) along the tubes ---1D optical lattices ($N=4$ of them are shown); Dashed arrows indicate the intra-plane $V_{xx'}$ and the inter-plane interactions $V_{zz'}$. }
\label{fig:setup}
\end{center}
\end{figure}

In the present work we discuss possible phases in a system of hardcore bosons confined to a stack of $N$ one-dimensional lattices---tubes (see Fig.~\ref{fig:setup}). Bosons in neighboring tubes interact via inter-tube interaction (either nearest-neighbor or dipole-dipole), with the inter-tube tunneling suppressed (this can be achieved experimentally with a deep optical lattice potential along the direction perpendicular to the tubes). 
Our focus is on quantum many-body phases of self-assembled chains of molecules from different tubes \cite{Wang,chains2,chains3,chains4,chains5,chains6,Potter2010}. 

In previous theoretical studies, mostly variational methods have been used. In Ref.~\cite{Capogrosso} the multilayered system has been mapped to a model amenable to classical Monte Carlo technique, and it has been shown that bosons in a stack of one-dimensional tubes can form superfluids of multi-atomic complexes -- chain superfluids (CSF) \cite{Capogrosso}, each chain consisting of one molecule from each tube, and there is, in general, a threshold for the CSF formation. An interesting opportunity for the emergence of exotic parafermions, as a generalization of Majorana fermions, in layered systems has been proposed in Refs.~\cite{Kuklov, Nonne} and tested by Monte Carlo simulations \cite{Kuklov}. It was also suggested that in two parallel 1D lattices with no inter-tube tunneling an exotic superfluid, consisting of $p\geq 1$ molecules from one tube and $q > 1$ ones form the other, should also be possible to realize \cite{Burovski,Demler_2011}. 

In contrast to what was previously done, here we study the actual quantum Hamiltonian of hard-core bosons by means of the {\it ab initio} path integral Quantum Monte Carlo (QMC) simulations (in continuous time) using a multi-worm algorithm \cite{Capogrosso} -- an extension of the Worm Algorithm \cite{WA} and its two-worm modification \cite{two_worms}. As we mentioned above, our algorithm is equally relevant to atomic mixtures and coupled spin chains.
We will show that CSF and other phases can be induced by infinitesimally small inter-layer interaction. Our study is a first step toward {\it ab-initio} simulations of more involved cases including
spin ladders and polar molecules with inter-tube tunneling. Using this algorithm it should be possible to provide accurate recommendations for the experimental realizations of the complex dipolar phases. 
\section{Hamiltonian}
The system under consideration is described by the single-band tight-binding Hamiltonian
\begin{widetext}
\begin{equation}
\label{eq:H}
 H= -J \sum_{<x,x'>,z}a_{xz}^\dagger a_{x'z}+ \frac{1}{2}\sum_{xz;x'z'} V(x-x',z-z') n_{xz}n_{x'z'}
-\sum_{xz}\mu_{z} \; n_{xz}
\end{equation}
\end{widetext}
 in grand canonical ensemble. Here $J>0$ stands for the intra-tube tunneling amplitude, $a^\dagger_{xz}$ ($a_{xz}$) is the creation (annihilation) operator for a hard core boson at site $(x,z)$, where $z=0,1,2, ...,N-1$ labels the tubes and $x=0,1,2,...,L$ is the coordinate along a tube. Here, $<>$ denotes summation over nearest neighbors, $ n_{xz}= a^\dagger_{xz}a_{xz}$, and $\mu_{z}$ is the chemical potential, which can be different in different tubes. 

The interaction $V(x,z)$ can be arbitrary. In the case of the dipole-dipole interaction, with the polarization axis being perpendicular to the tubes and belonging to the plane of the tubes, it takes the form
\begin{equation}
\label{eq:Vd}
 V(x,z)= V_d \frac{x^2 - 2z^2}{(x^2 +z^2)^{5/2}},
\end{equation}
where $V_d>0$ sets the energy scale. In this geometry, the interaction along the $z$-axis is attractive. As we will discuss below, in 1D, {\it arbitrary small} $V_d$ can induce superfluidity of quasi-molecular complexes. This result follows from the bosonization analysis and has previously been noted for the case of pairing of hard core bosons in Ref.~\cite{Mathey}.

The repulsive part of the interaction along the $x$-axis favors solidification.
A special role is played by the filling factor $\nu=1/2$. As we will show later, in the case $N>2$ the insulating phase featuring 1D checkerboard order emerges in the limit $V_d \to 0$  even if no intra-layer repulsion is explicitly introduced. 

When dipoles are polarized perpendicularly to the tubes plane, the interaction becomes purely repulsive,
\begin{equation}
\label{eq:Vdrep}
 V(x,z)= V_d \frac{1}{(x^2 +z^2)^{3/2}},
\end{equation}
and can result in super-counterfluid (SCF) \cite{SCF} phases which are also thresholdless with respect to the interaction.

\section{Density controlled quantum phases in layered systems} \label{linear_response}
A system of hard-core bosons, trapped in one-dimensional tubes with no inter-tube Josephson coupling, forms $N$ independent superfluids characterized by $N$  quasi-condensate order parameters $\langle\psi_z \rangle\sim \exp(i\phi_z)$ with phases $\phi_z, \, z=0,1,2,...N-1$. The hard-core nature of bosons in each tube plays a special role. As we will see below, an arbitrary small inter-tube interaction can induce multiplicity of various superfluid and insulating phases depending on the filling factors $\nu_z$ in the tubes.  The counter intuitive threshold-less nature of the phases simply means that observing them is possible for arbitrary small $V_d$ on correspondingly large spatial scales. It is worth noting that, depending on a combination of the filling factors, various types of mixtures of such phases can exist as well.  

\subsection{Thouless phase twists and windings}
Here we introduce a description in terms of the generalized superfluid stiffness $R_{zz'}$ and superfluid compressibility $C_{zz'}$.
This language of the generalized superfluid response turns out to be very helpful in defining  ground states of the bosonic complexes as well as in characterizing  ground states numerically. 
The response matrices are defined through contributions to the system action as a result of imposing infinitesimal Thouless phase twists $\vec{\phi}'(z)=(\phi'_x(z), \phi'_\tau(z))$ on the space-time boundaries of the tubes. Such twists can be viewed in terms of the corresponding gauge potentials $A_x(z,x) = \phi'_x(z)/L$ along space and $A_\tau(z,\tau) = \phi'_\tau(z)/\beta$ along time, where $L,\;\beta$ stand for tubes length and inverse temperature in atomic units, respectively. It is important that, in the case of the periodic boundary conditions on the phases $\phi_z(x,\tau)$ of the fields%{\color{red}can't we just write PBC on fields?$\Psi_{i+L}=\Psi_i$}
, such gauge potentials cannot be absorbed into the phases.

In general, the infinitesimal contribution of the twists to the action is given by:
\be
\label{TH}
E= \sum_{zz'}\left[\frac{\beta}{2L} R_{zz'} \phi'_x(z)\phi'_x(z') +\frac{L}{2\beta} C_{zz'} \phi'_\tau(z)\phi'_\tau(z')\right] .
\ee
The quantities $R_{zz'}$ and $C_{zz'}$ can be expressed in terms of topological properties of the particle world-lines, windings $\vec{W}(z)=(W_x(z), W_\tau(z)) $, and can be measured numerically. Global gauge invariance of the system implies that the total partition function $Z=Tr(\exp(-\beta H))$ 
can be represented as a statistical sum over all possible winding numbers of closed world lines of particles as
\be
\label{Z}
Z= \sum_{\{\vec{W}(z)\}}Z[\{\vec{W}\}] \exp[ i \sum_{z} \vec{W}(z)\vec{\phi}'(z)] ,
\ee 
where $Z[\{\vec{W}\}]$ stands for a functional of windings in all tubes. The superfluid stiffnesses can be obtained as second derivatives of $E= -\ln Z$ with respect to $(\phi'_x(z),\phi'_\tau(z))$ in the limit $\vec{\phi}'(z) \to 0$ as 
\be
\label{R}
R_{zz'}= \frac{L}{\beta} [\langle W_x(z) W_x(z')\rangle - \langle W_x(z)\rangle \langle W_x(z')\rangle], 
\ee 
\be
\label{C}
C_{zz'}= \frac{\beta}{L} [\langle W_\tau(z) W_\tau(z')\rangle - \langle W_\tau(z)\rangle \langle W_\tau(z')\rangle]. 
\ee 
As long as the tubes are identical, $R_{zz'}$ and $C_{zz'}$ depend on the difference $z-z'$, $R_{zz'}=R(z-z')$ and $C_{zz'}=C(z-z')$. Hence, the Fourier transform along $z$-axis can be used, $\tilde{R}(q_z) = \sum_z R(z) \exp( iq_z z), \, \tilde{C}(q_z) = \sum_z C(z) \exp( iq_z z)$, where $q_z= 2\pi m_z/N, \, m_z=0,1,2,...,N-1$. Then, Eq.~\eqref{Z}  expressed in terms of the Fourier transforms $ \tilde{\phi}'_{x,\tau}(q_z) =N^{-1/2} \sum_z \phi'_{x,\tau}(z) \exp( iq_z z),\, \tilde{W}_{x,\tau}(q_z) =N^{-1/2} \sum_z W_{x,\tau}(z) \exp( iq_z z)$ gives
\be
\label{FurR}
\tilde{R}(q_z)= \frac{L}{\beta} [\langle \tilde{W}_x(q_z)\tilde{W}_x(-q_z)\rangle- \langle \tilde{W}_x(q_z)\rangle \langle\tilde{W}_x(-q_z)\rangle]
\ee
and
\be
\label{FurC}
\tilde{C}(q_z)= \frac{\beta}{L} [\langle \tilde{W}_\tau(q_z)\tilde{W}_\tau(-q_z)\rangle- \langle \tilde{W}_\tau(q_z)\rangle \langle\tilde{W}_\tau(-q_z)\rangle]
\ee 
These equations represent an extension of the Ceperley and Pollock expression \cite{Ceperley} for the superfluid stiffness and compressibility.

In full analogy with the case $N=1$, the ratio $V_s(q_z)=\sqrt{\tilde R(q_z) /\tilde{C}(q_z)} $ has the meaning of the speed of sound propagating along tubes with dispersion along the $z$-axis. Extending the analogy, the product $\tilde{R}(q_z) \tilde{C}(q_z)$ gives the Luttinger ``parameter" (rather, Luttinger matrix) as
\be
\label{Lut}
\tilde{K}(q_z) = \pi \sqrt{\tilde{R}(q_z) \tilde{C}(q_z)}.
\ee
Thus, the action for arbitrary (small) phase fluctuations  of the translationally invariant (along $z$-axis) system  renormalized by the interactions becomes 
\begin{widetext}
\be
\label{Full}
S_R= \int_0^\beta d\tau \int dx \sum_{q_z} \left[ \frac{V_s(q_z)\tilde{K}(q_z)}{2\pi} |\nabla_x \tilde{\phi}_{q_z}(x,\tau)|^2 +\frac{\tilde{K}(q_z)}{2\pi V_s(q_z)} |\nabla_\tau \tilde{\phi}_{q_z}(x,\tau)|^2 \right ],
\ee 
\end{widetext}
where $\tilde{\phi}_{q_z}(x,\tau)$ are the Fourier components of the phases $\phi_z(x,\tau)$ with respect to the tube index $z$.

The speed of sound is not significantly renormalized compared to the strong renormalization of superfluid stiffness $R$ and compressibility $C$~\cite{Mathey}. This simply means that the space-time symmetry of the superfluid-insulator transitions is preserved in translationally invariant system. Thus, for all practical purposes, the dispersion of the speed of sound $V_s$ vs $q_z$ can be ignored so that $R_{zz'} = C_{zz'}$ in units of $V_s=1$. In this limit, the Luttinger matrix and the matrix of stiffnesses are equivalent to each other.
Then, the generalized linear response can be fully described by the following translationally invariant action $S_R = \int_0^\beta d\tau \int dx \sum_{q_z} \frac{1}{2\pi }\tilde{K}(q_z) |\vec{\nabla}\tilde{\phi}(q_z)|^2$, or in the direct z-space as
\begin{equation}
\label{Rphi}
 S_R = \int_0^\beta d\tau \int dx \sum_{z,z'} \frac{1}{2\pi }( \hat{K})_{z,z'} \vec{\nabla}\phi_z\vec{\nabla}\phi_{z'},
\end{equation}
where $\hat{K}$ stands for Luttinger parameter matrix with the dimension $N \times N$. It is worth mentioning that this form features the non-viscous drag between superfluid flows in different tubes. It is responsible for the formation of the complex superfluid and supercounterfluid phases.  We will be referring to the form (\ref{Rphi}) and, specifically, to the properties of the Luttinger matrix $\hat{K}$ while identifying the ground states of the bosonic complexes. 

\subsection{N atomic superfluids}\label{AS}
If the matrix $\hat{K}$ in (\ref{Rphi}) is non-degenerate (in the case when all filling factors are different and not complimentary to unity), there exists the standard algebraic (or 1D superfluid) order  in the correlators $\langle \exp(i\phi_z(x)) \exp(- i\phi_z(0))\rangle \sim 1/|x|^{(\hat{K}^{-1})_{zz}/2}$ (and $\langle \exp(i\phi_z(x)) \exp(- i\phi_{z'}(0))\rangle =0$ for $z\neq z'$). A sketch of this phase is shown in Fig.~\ref{fig:NSF}. As we will see below, should some filling factors become the same or complimentary to unity, the inter-tube interaction can easily destroy such atomic orders in favor of composite superfluids or supercounterfluids.
\begin{figure}[h]
\begin{center}
\includegraphics[width=0.3\textwidth]{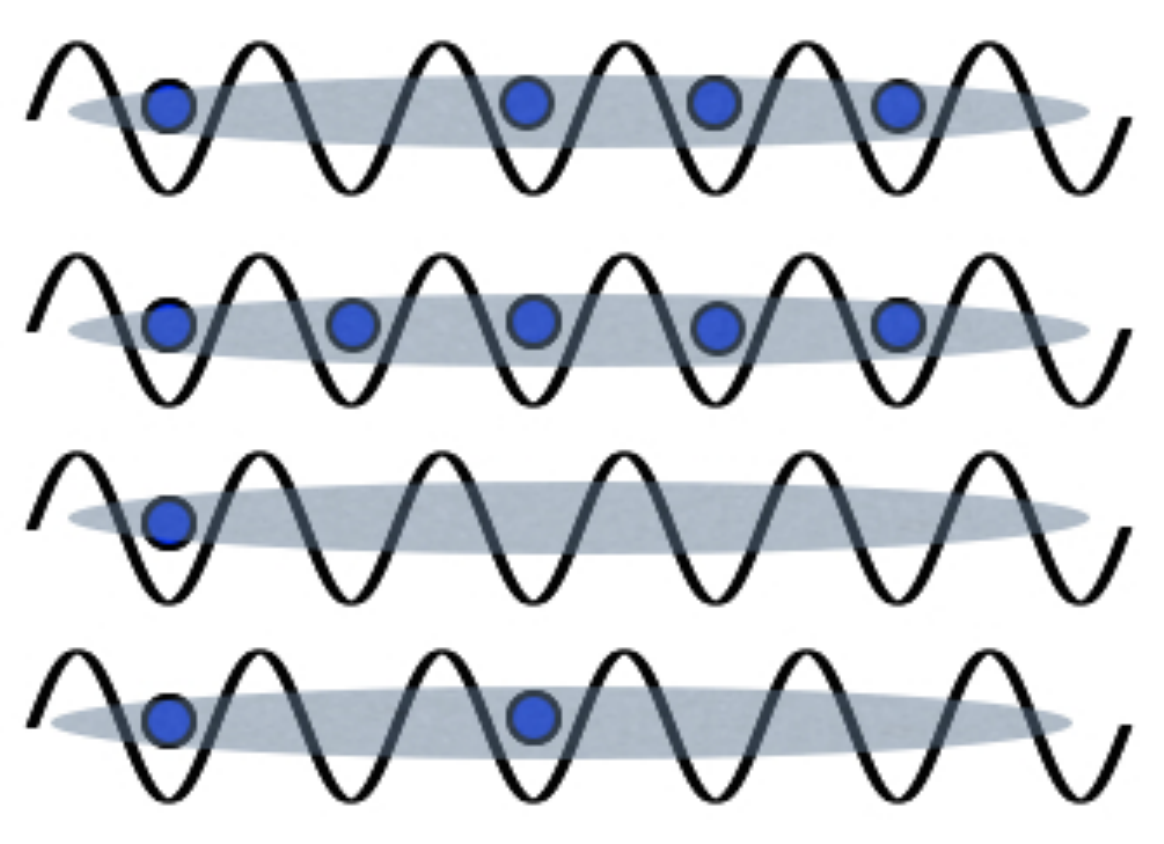}
\caption{A sketch of the N-atomic superfluids characterized by N independent algebraic orders (depicted by N=4 fuzzy clouds).}
\label{fig:NSF}
\end{center}
\end{figure}

\subsection{Composite superfluids}\label{CS}
\begin{figure}[h]
\begin{center}
\includegraphics[width=0.3\textwidth]{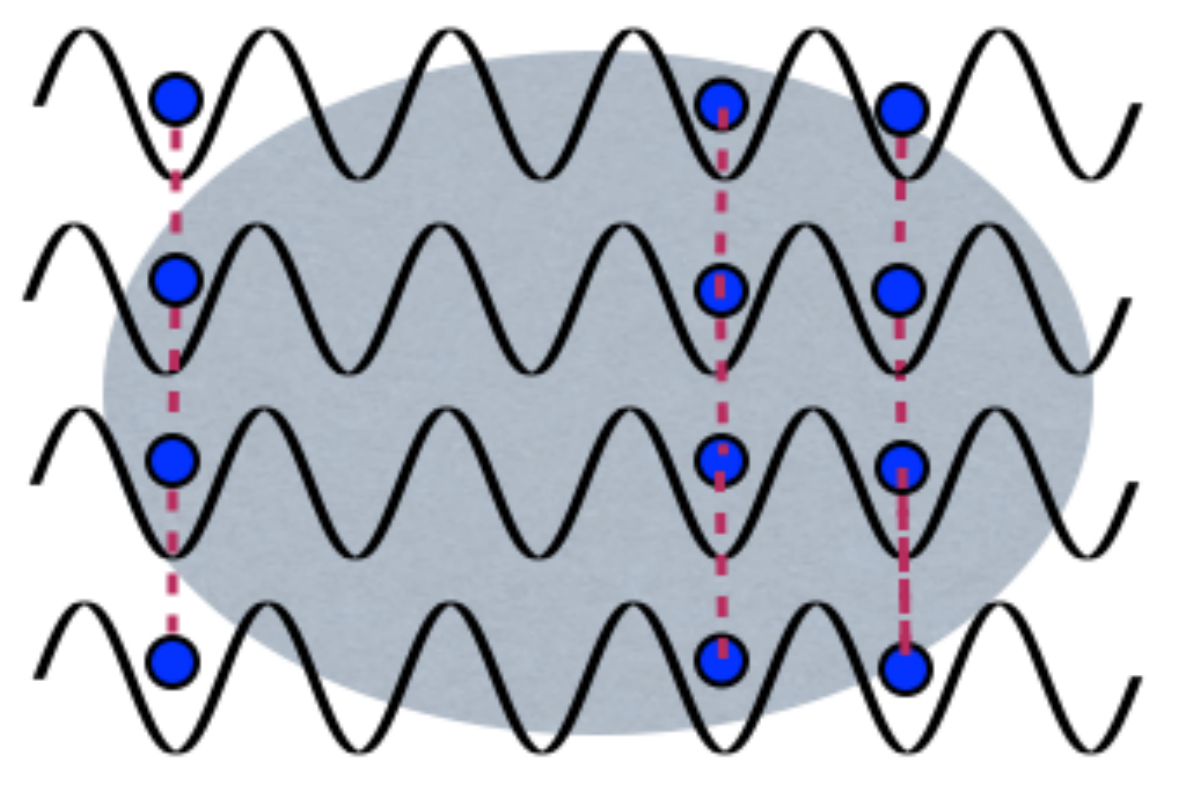}
\caption{A sketch of the CSF phase.  Dashed lines indicate binding of atoms from different tubes and the fuzzy cloud depicts single algebraic order for all tubes.}
\label{fig:CSF}
\end{center}
\end{figure}

If all $N$ tubes are characterized by the same incommensurate filling factor $\nu$, the nature of the superfluid correlations changes dramatically as long as there is an arbitrary small attraction between the tubes. Specifically, the matrix $\hat{K}$ becomes degenerate so that $\langle \exp(i\phi_z(x)) \exp(- i\phi_z(0))\rangle $ decays exponentially. The algebraic decay will be observed only in the $N$-body density matrix $\langle \Psi^\dagger(x) \Psi(0)\rangle$ where $\Psi(x) =  \psi_{z=0}(x)\psi_{z=1}(x)...\psi_{z=N}(x)$.
In other words, the matrix elements of $\hat{K}$ in Eq.(\ref{Rphi})  become all identical to each other, so that $N-1$ eigenvalues are equal to zero and only one remains finite --- corresponding to a finite superfluid stiffness of the sum of the phases $\Phi= \sum_z \phi_z$. In terms of the Fourier components of the matrix kernel, $\tilde{K}(q_z=0) \neq 0$ while $\tilde{K}(q_z\neq 0) = 0$.  This defines the CSF, a superfluid of quasi-molecular complexes, each complex consisting of $N$ bosons -- one from   
each tube. A sketch of this phase is shown in Fig.~\ref{fig:CSF}.

If $N>2$, it is possible to have a situation when only $1<M<N$ tubes have identical filling factors. Then, the composite superfluid will be formed among these tubes while others carry the standard atomic superfluids. In general, a group of $M$ layers with the same filling factor adds degree of degeneracy M-1 to the matrix $\hat{K}$. In other words, the  number of the remaining superfluid phases is equal to $N$ minus the total degree of degeneracy. This means that the  matrix $\hat{K}$ will have as many zero eigenvalues as there exist
restored U(1) symmetries. As we will discuss below, such phases can be realized for arbitrary small inter-tube interaction $V_d$.

 \subsection{Supercounterfluids}\label{SCF}
\begin{figure}[h]
\begin{center}
\includegraphics[width=0.3\textwidth]{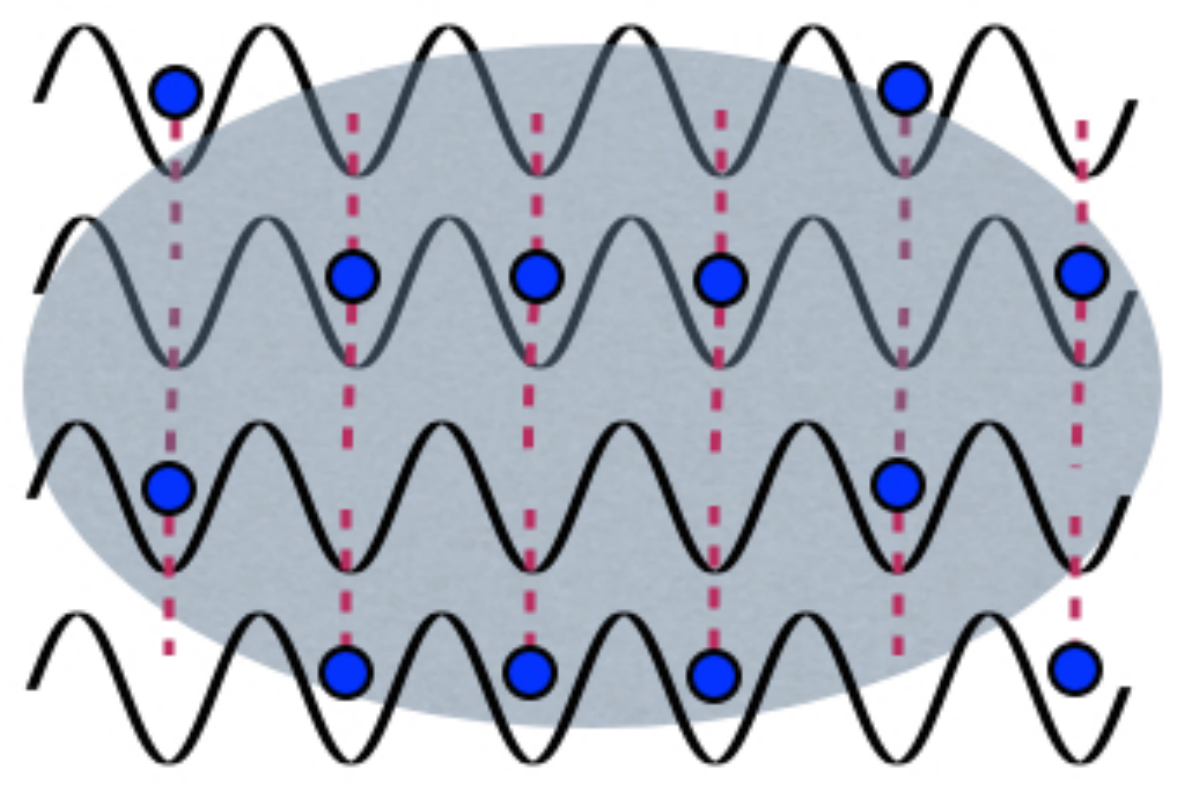}
\caption{A sketch of the SCF phase in  $N>2$ tubes. Similarly to Fig.~\ref{fig:CSF}, the fuzzy cloud depicts a single algebraic order for all tubes. The dashed lines indicate binding between atoms and holes in the tubes with complementary fillings and between atoms in the tubes with the same fillings.}
\label{fig:SCF}
\end{center}
\end{figure}

The concept of supercounterfluidity (SCF) has been introduced for two-component systems in Ref.\cite{SCF}. SCF can exist in a lattice when the filling factors $\nu_1$ and $\nu_2$ for both components complement each other to an integer filling, $\nu_1 + \nu_2=1$. Then, the repulsive interaction can induce binding of atoms of sort ``1" to holes of sort ``2". Using the language of broken symmetries, the U(1)$\times$ U(1) symmetry becomes partially restored so that only one U(1) symmetry remains broken. In terms of fields, the field $\exp(i\phi_1)\exp(-i\phi_2)$ is condensed while $\exp(i\phi_1)\exp(i\phi_2)$ becomes disordered. Accordingly, the superflow can only exist in the counterflow manner -- when transfer of one atom of sort ``1" is compensated by motion of one atom of sort ``2" in the opposite direction.  This property can naturally be extended to a general case of $N$ sorts of atoms when the superflow of, say, $M<N$ components  is (partially) compensated by the counter-flow of the remaining components. The SCF phase is sketched in Fig.~\ref{fig:SCF}.

In general, there could be $M_1$ tubes all with the same filling factors $\nu_1$ and $M_2$ tubes also with identical filling factors  $\nu_2=1-\nu_1$ so that $\nu_2 \neq \nu_1$. Thus, there are two groups of the composite superfluids, consisting of $M_1$ and $M_2$ complexes. Accordingly, there are $M_1-1 + M_2-1$ restored symmetries. Moreover, the backscattering (BS) interaction between particles from the first and the second groups restores one additional symmetry. The corresponding composite operator which characterizes the algebraic order is    $\Phi_{M_1,M_2}(x)= \psi_{z_1}(x)...\psi_{z_{M_{1}}}(x)\psi^\dagger_{z'_1}(x)...\psi^\dagger_{z'_{M_2}}(x)$, where $z_1, z_2,...,z_{M_1}$ label tubes from the first group and $z'_1,z'_2,...,z'_{M_2}$ --- from the second. In such a phase a transfer of $M_1$ atoms from the first group is compensated by the counter-motion of $M_2$ atoms from the second group, so that there is a net transfer of $M_1 -M_2$ atoms.
Accordingly, $M_1 +M_2 -1$ eigenvalues of the matrix $\hat{K}$ in Eq.(\ref{Rphi}) are zero. In other words, the resulting state can be thought of as a bound state of two composite superfluids in the counter-flow regime --- a natural generalization of the two-component SCF \cite{SCF}.

In the special case when all, e.g , odd tubes have filling factor $\nu$ and all even ones have $1-\nu$ (as exemplified in Fig.~\ref{fig:SCF}), the Fourier transform can be used. In this case all Fourier harmonics but $\tilde{K}(q_z=\pi)$ are equal to zero, so that there is no net transfer of atoms. As we will discuss below, such a phase can also be realized for arbitrary small $V_d$.

\subsection{Composite insulators}\label{CI} 
\begin{figure}[h]
\begin{center}
\includegraphics[width=0.3\textwidth]{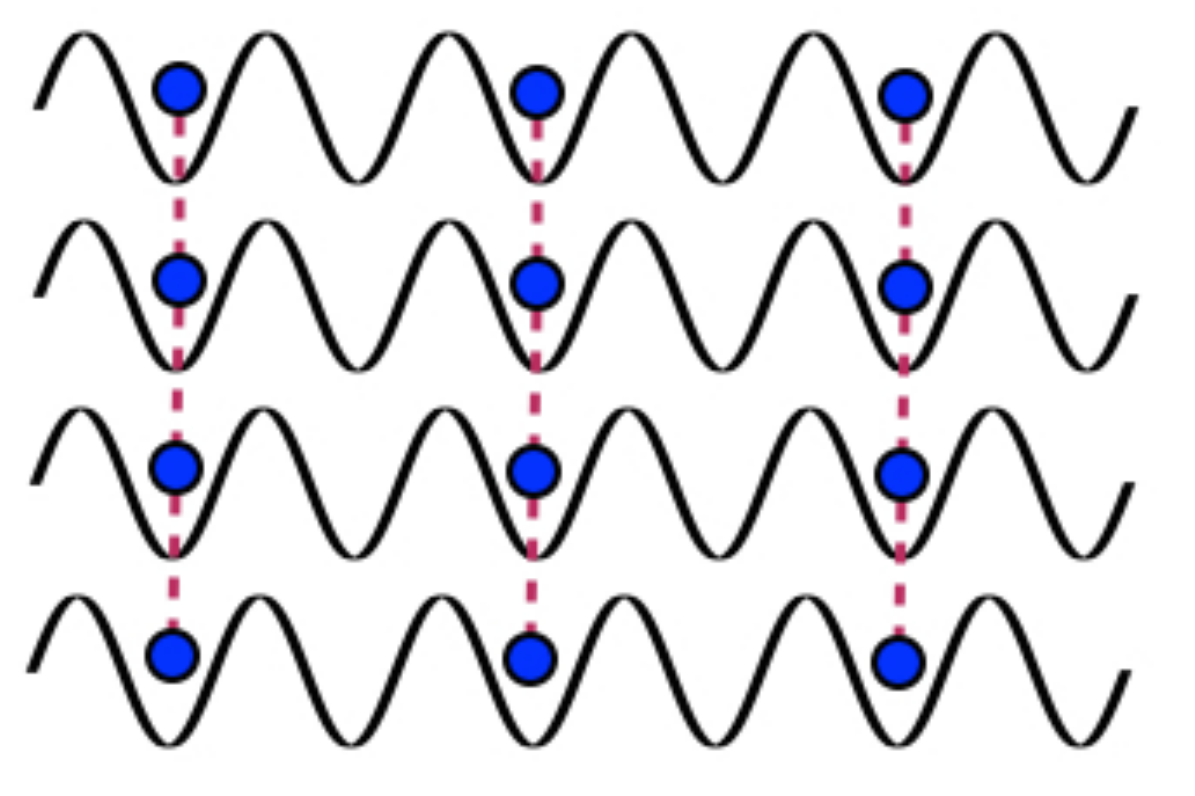}
\caption{A sketch of the CB phase. The dashed lines indicate binding between atoms from different tubes at $\nu=1/2$.}
\label{fig:CB}
\end{center}
\end{figure}

The easiest way to form an insulator in 1D lattices is at filling factor $\nu=1/2$. In a single tube ($N=1$) at $\nu=1/2$ the checkerboard (CB) type solid can exist only if the two-body repulsion exceeds a certain threshold. The situation becomes dramatically different in the cases  $N > 2$. As it will be discussed below, the bosonization analysis shows that, even in the absence of any repulsion, the inter-layer attraction induces the  CB insulator in the limit $V_d \to 0$ as long as $N\geq 3$. This conclusion is consistent with our {\it ab initio} simulations. A sketch of this phase is shown in Fig.~\ref{fig:CB}.

%The situation with $N=2$ is not that conclusive. While the bosonization predicts no insulating state in the absence of the repulsion and, instead, a superfluidity of pairs with the Luttinger parameter $K>1$, the simulations show that $K$ flows toward a superfluid fixed point where $K\approx 0.5$, that is, to a state which should be insulating according to the bosonization.  At the moment we don't have an explanation to this feature.

 Solids at other rational fillings $\nu= 1/3, 1/5,2/5, ...$ are possible as well. To induce them, however, the  interaction $\sim V_d$ must exceed the corresponding thresholds determined by the denominators of the fractions. 
Concluding this section we note that in an insulating state the renormalized Luttinger matrix $\hat{K}$ in the action~\eqref{Rphi} is zero. 

\section{N-tube bosonization } 
Here we will discuss the phases described in \ref{AS}, \ref{CS}, \ref{SCF}, \ref{CI} within the framework of the bosonization approach \cite{Haldane_1980} in order to reveal their threshold-less nature.
The bosonic field operator $\psi_z(x)$ is represented in terms of the superfluid phase $\phi_z$ and the density $\rho_z =|\psi_z|^2$, which can be expressed in terms of  Haldane's ``angle" variable $\theta_z$ \cite{Haldane_1980} as  
\begin{equation}
\label{eq:rho}
\rho_z(x)=(\nu_z+\frac{1}{\pi}\nabla_x\theta_z) \sum_{m_z=0, \pm 1, ...} e^{\left[2m_z i(\theta_z+\pi \nu_z x)\right]}\,,
\end{equation}
where $\theta_z(x)$ is conjugate to the superfluid phase $\phi_z$. The term $m_z=0$ gives the forward scattering (FS) interaction and the terms with $m_z \neq 0$ account for the back scattering (BS) events. 

In the absence of inter-tube tunneling, the bosonized action corresponding to the Hamiltonian \eqref{eq:H} is 
\begin{equation}
\label{Stotal}
S=\sum_z S^{(0)}_z + \sum_{z,z'} [ S^{\rm (FS)}_{z,z'} + S^{\rm (BS)}_{z,z'}],
\end{equation}
 where 
\begin{align}
\label{Sz}
\nonumber S^{(0)}_z = \int_0^\beta d\tau \int dx \frac{1}{2\pi K} &[ (\nabla_\tau \theta_z(x,\tau))^2\\
&+(\nabla_x\theta_z(x,\tau))^2],
\end{align}
in units $V_s=1$. Here $K$ is the bare Luttinger liquid parameter, that is, not yet renormalized by the interactions. For hard-core bosons and zero dipolar interaction, each tube is equivalent to a XXZ spin $S=1/2 $ chain with zero $S_z$-$S_z$ coupling. Accordingly, $K=1$ (see in Ref. \cite{Giamarchi}). In the following we use periodic boundary conditions along $x$- and $z$-coordinates.

The second term in Eq.~\eqref{Stotal} is the FS part of the action, 
%\begin{multline}
%\label{S_FS1}
$ S^{\rm (FS)}_{z,z'} = \frac{1}{2 \pi^2} \int_0^\beta d\tau \int dx \int dx' V(x-x',z-z')  % \\
\nabla_x \theta_z(x,\tau)\nabla_{x'} \theta_{z'}(x',\tau)$.
%\end{multline}
In the long wave limit of the space-time Fourier representation ($q_x\rightarrow 0$) the FS part of the action becomes 
% S^{\rm (FS)}_{z,z'} = \frac{1}{2} \sum_{\vec{q},q_z} \tilde{V}(q_z)q^2_x |\tilde{\theta}(q_z,\vec{q})|^2 .
\begin{equation}
\label{S_FS2}
S^{\rm (FS)}_{z,z'} = \frac{1}{2} \sum_{\vec{q}} \bar{V}(z-z')q^2_x \tilde{\theta}_z(\vec{q})\tilde{\theta}_{z'}(-\vec{q})\;.
\end{equation}
Here, the summation is performed over the time-space harmonics, $\vec{q}= (\omega, q_x)$ along each tube %as well as along the discrete Fourier vectors $q_z = 2\pi n_z/N,\, n_z=0,1, ...,N-1$ along $z$-axis, so that $\theta_z(x,\tau) = \frac{1}{\sqrt{N \beta L}} \sum_{\vec{q},q_z} \exp[ i \omega \tau + iq_x x + i q_z z] \tilde{\theta}(q_z,\vec{q}) $; 
and 
\begin{equation}
\label{Vtil} 
\bar{V}(z)=\frac{1}{\pi^2} \int dx V(x,z).
 \end{equation}
In particular, for the dipole-dipole interaction given by Eq.~\eqref{eq:Vd} one finds
\begin{equation}
\label{Vdip} 
\bar{V}(z-z')= - \frac{\gamma_1}{(z-z')^2}, \quad \gamma_1 \approx \frac{2.00 V_d}{\pi^2}
 \end{equation}
for $|z-z'|=1,2, 3,...$. For $z=z'$, the dipolar interaction is purely repulsive, with $\bar{V}(0) =\int dx V(x,z=0)$, and it must be cut off at some short distance. 
Here we consider the same length scale along $z$ and $x$, and choose the cutoff at $x=1$ such that $ \bar{V}(0) \approx 2.40 V_d/ \pi^2 $. Thus, in the long-wave limit, while the inter-layer interaction
is attractive, the intra-layer one is repulsive. In the case of the purely repulsive dipolar interaction, Eq.\eqref{eq:Vdrep}, the FS interaction given by Eq.~\eqref{Vdip} changes sign, that is, $\gamma_1 \to -\gamma_1$ for $z\neq z'$ (with $ \bar{V}(0)$ being unchanged).

Next, we introduce Fourier harmonics along the $z$-coordinate, $ \tilde{\theta}_{q_z}(\vec{q})$ and $\tilde{V}(q_z)$, corresponding to $\tilde{\theta}_{z}(\vec{q})$ and $\bar{V}(z)$, respectively. We can now write the Gaussian part  $\sum_z S^{(0)}_z + \sum_{z,z'} S^{(FS)}_{z,z'} $ of the action (\ref{Stotal}) as
\begin{equation}
\label{gauss}
 S_0 = \sum_{\vec{q},q_z} \left [ \frac{1}{2\pi K} \vec{q}\;^2 + \frac{1}{2}\tilde{V}(q_z)q^2_x\right]|\tilde{\theta}(\vec{q},q_z)|^2.
 %S_0 = \sum_{\vec{q},q_z} \frac{1}{2\pi K} \vec{q}\;^2 \vert \tilde{\theta}(\vec{q},q_z)\vert^2,
\end{equation}
Eq.~\eqref{gauss} implies the renormalization of the speed of sound $V_s \to \tilde{V}_s(q_z)=\sqrt{1+ \pi K \tilde{V}(q_z)}$ (in units of the bare value) as well as of the Luttinger parameter 
\begin{equation}
\label{K}
K \to \tilde{K}(q_z)= \frac{K}{\sqrt{ 1+ \pi K \tilde{V}(q_z)}}\;.
\end{equation}
Thus, both quantities $\tilde{V}_s(q_z),\,\tilde{K}(q_z)$ depend on the wave-vector $q_z$ counting the layers so that the action (\ref{gauss}) takes the form (\ref{Full}). As discussed above, in what follows we will ignore the renormalization of the speed of sound and will rather consider the form (\ref{Rphi}). Then, in terms of the dual variables $\theta_z$ the gradient part of the renormalized action becomes
\begin{equation}
\label{Rgauss}
 S_R = \int dx \int_0^\beta d\tau \sum_{z,z'} \frac{1}{2\pi }( \hat{K}^{-1})_{z,z'} \vec{\nabla}\theta_z\vec{\nabla}\theta_{z'}, 
\end{equation}
where $(\hat{K}^{-1})_{z,z'}$ is the inverse of the renormalized Luttinger matrix $\hat{K}$ introduced in Eq.~\eqref{Rphi}.

Finally, the third term in Eq.(\ref{Stotal}) accounts for the backscattering events \cite{Haldane_1980} which in the context of the system studied can be written as
\begin{widetext}
\begin{equation}
\label{BS}
S^{\rm (BS)}_{z,z'} = - \int d\tau \sum_{m_z,m_{z'}}\sum_{x=0,1,...L} V_{m_z, m_{z'}}(z,z')\cos[ 2 ( m_z \theta_z + m_{z'} \theta_{z'} )
 + 2\pi (\nu_z m_z + \nu_{z'} m_{z'}) x]\;, 
\end{equation}
\end{widetext}
where the amplitudes $ V_{m_z; m'_{z'}}(z,z')$ are induced by the interaction and satisfy the renormalization flow (to be derived in the standard one-loop approximation in Appendix A). 
While the FS sets in the initial value of the Luttinger matrix Eq.\eqref{K}, the BS is responsible for its further renormalization.

In the limit $V_d \to 0$ the renormalization group (RG) flows can be found exactly. First we remind that for hard-core bosons $K_{zz}=1$ and $ K_{zz'}=0$ for $z\neq z'$ at $V_d=0$. Therefore, as it will be clear below, in the limit of small interactions, only the lowest non-trivial values of $m_z, m_{z'}$ ($m_z = \pm 1, m_{z'} = \pm 1$) can become relevant in the sum (\ref{BS}), provided $\nu_z m_z + \nu_{z'} m_{z'} =0, \pm 1$. Hence, the relevance of the backscattering for a particular pair $ z, z'$ of layers can be controlled by adjusting the bosonic populations in individual tubes. 
%%%%%%%%%%%%%%%%
\subsection{RG for the composite superfluid}
Due to the spatially non-local nature of dipolar interactions, the composite superfluid phase, \ref{CS}, can form between tubes with the same filling factors regardless of their geometrical positions. For example, in a system of $N=6$ tubes where $\nu_1= \nu_2=\nu_5=\nu$ (here we consider $\nu \neq 1/2$) , with all other values $\nu_3\neq  \nu_4 \neq  \nu_6\neq \nu$, the harmonics     $V_{1;-1}(1,2), V_{1;-1}(1,5), V_{1;-1}(2,5)$ can  become relevant, while all others remain irrelevant (simply because of the oscillating phases $ 2\pi (\nu_z m_z + \nu_{z'} m_{z'}) x$, with $z,z' =3,4,6$, in the corresponding cos-harmonics in Eq.(\ref{BS})).  

 The RG equations for the amplitudes of the corresponding harmonics ($z\neq z'$) between the tubes with identical filling factors are (see Appendix \ref{A}) 
\begin{equation}
\label{RG11}
\frac{dV_{1;-1}(z,z')}{d\ln l}= \left[2- K_{zz} - K_{z'z'} +2K_{zz'} \right] V_{1;-1}(z,z'),
\end{equation}
where $K_{zz'}$ are the matrix elements of the matrix $\hat{K}$ in Eq.(\ref{Rphi}) and the initial (bare) values of the amplitudes $V_{1;-1}(z,z')$ are determined by the dipolar interactions. In the limit of no interactions, that is $\bar{V} \to 0$, the RG flow starts from the critical point determined by the factor  $\left[2- K_{zz} - K_{z'z'} +2K_{zz'} \right] \to 0$ (since  $K_{zz'} \to K \delta_{zz'}$, with $K=1$) in Eq.(\ref{RG11}). As explained in the Appendix A, Eq.(\ref{RG22}), this factor is $\propto V_d$ and is positive for the case of the attractive inter-layer interaction  (\ref{Vdip}). Thus, the relevant amplitudes diverge as $V_{1;-1}(z,z') \sim l^b$ with $ b \sim V_d$. While formally this implies that CSF is induced by an arbitrary small interlayer attraction $V_d$, a physical scale $ l_{CSF}$ on which such a composite phase can be observed is actually exponentially divergent as  $l_{\rm CSF}\sim \exp(... 1/V_d) \to \infty $, where "..." means a coefficient $\sim 1$ (see below).  

Using the $N=6$ example from above, the formation of the CSF between the tubes  $z=1,2,5$ implies that the harmonics $V_{1;-1}(1,2), V_{1;-1}(1,5), V_{1;-1}(2,5)$ exhibit the runaway flow to $\infty $ in Eqs.(\ref{RG11}), while all other combinations can be set essentially to zero. In other words, two U(1) symmetries are being restored so that the system is decsribed by the algebraic orders in $\psi_3, \psi_4, \psi_6$ and in the CSF field $\Psi_{1,2,5} = \psi_1 \psi_2 \psi_5$.

In the case of translational invariance along the $z$-axis, that is, when $\nu_z=\nu$ for all tubes, Eq.(\ref{RG11})  can be explicitly written in terms of the kernels of Luttinger matrix and its inverse as 
\begin{equation}
\label{RG111M}
\frac{dV_{1;-1}(z)}{d\ln l}=2 \left[1- K(0) + K(z)\right] V_{1;-1}(z),
\end{equation} 
%\begin{equation}
%\label{RG111M}
%\frac{dV_{1;-1}(z)}{d\ln l}= \left[2- \frac{4}{N} \sum_{q_z} \tilde{K}(q_z) \sin^2\left(\frac{q_zz}{2}\right)\right] V_{1;-1}(z),
%\end{equation} 
where we have taken into account that the amplitudes $V_{1;-1}(z,z')$ as well as the matrix elements $K_{zz'}$ are functions of the difference $z-z'$ rather than of $z,z'$ separately: $ V_{1;-1}(z,z') \equiv  V_{1;-1}(z-z'),\,(\hat{ K})_{zz'}\equiv K(z-z'),\,(\hat{ K}^{-1})_{zz'}\equiv K^{-1}(z-z')   $, where
\begin{eqnarray}
\label{FurM}
K(z)= \frac{1}{N} \sum_{q_z} \tilde{K}(q_z){\rm e}^{iq_z z},
\\
\label{Fu2rM}
K^{-1}(z)= \frac{1}{N} \sum_{q_z} \frac{1}{\tilde{K}(q_z)}{\rm e}^{iq_z z},
\end{eqnarray}
with the corresponding inverse transformations.

Ignoring the renormalization of the Luttinger matrix by the BS, the value of $\tilde{K}(q_z)$ from Eq.~\eqref{K} can be used in Eq.~\eqref{RG111M} in the limit $V_d \to 0$. Then, in the lowest order in $V_d$ we find 
\begin{equation}
\label{RG22M}
\frac{dV_{1;-1}(z)}{d\ln l}\approx  \pi (\bar{V}(0)- \bar{V}(z)) V_{1;-1}(z),
\end{equation}
where $\bar{V}(z)$ is given in Eq.~\eqref{Vdip}) and $K$ is set to its value, $K=1$, for non-interacting tubes. Thus, for $\bar{V}(0) >0$ and $\bar{V}(z-z') <0$, as it is in the case of the dipolar interaction between molecules polarized along the z-axis, Eq.(\ref{Vdip}), the harmonics $V_{1;-1}(z-z') \cos(2\theta_z - 2\theta_{z'})$ become relevant for arbitrary small $V_d$. This implies that the superflow is only possible in the channel of the center of mass motion of all tubes because relative density fluctuations are gapped.
It is also interesting to note that, in the case of the purely repulsive interaction (\ref{eq:Vdrep}) (that is, when the molecules are polarized along y-axis), where $\bar{V}(z) >0$ for $|z|>0$, the composite superfluid, CSF, is also possible as long as $\bar{V}(0) > \bar{V}(z) >0$. This binding caused by repulsion is a specific property of 1D geometry. 

The renormalization of the BS amplitudes, Eq.~\eqref{RG111M}, should be considered together with the renormalization of the matrix $\hat{K}$ in Eq.(\ref{Rphi}). 
As explained in the Appendix \ref{A}, these equations are 
\begin{equation}
\label{RG2M}
\frac{d(\hat{K}^{-1})_{zz'}}{d\ln l} =-C [V_{1,-1}(z,z')]^2(K_{zz} + K_{z'z'} - 2K_{zz'})
\end{equation}
for the off-diagonal terms, $z\neq z'$, and
\begin{equation}
\label{RG2dM}
\frac{d(\hat{K}^{-1})_{zz}}{d\ln l} =C \sum_{z'} [V_{1,-1}(z,z')]^2(K_{zz} + K_{z'z'} - 2K_{zz'})
\end{equation}
for the diagonal ones. Here the constant $C>0$ depends on the type of the short-distance cutoff (see in Ref.~\cite{Giamarchi}). This constant can be absorbed into the
definition of $V_{1,-1}(z)$ by simple rescaling of the amplitudes.
It is worth noting that only the pairs $(z,z')$ such that $\nu_z=\nu_{z'}$ are involved in Eq.~\eqref{RG2M}) and Eq.~\eqref{RG2dM}). 

In the case of the translational invariance, that is, $\nu_z=\nu$, these equations become 
\begin{equation}
\label{RG2FM}
\frac{dK^{-1}(z)}{d\ln l} =-2C[V_{1,-1}(z)]^2\left(K(0) - K(z)\right),
\end{equation}
where $z\neq 0$ and
\begin{equation}
\label{RG0M}
\frac{dK^{-1}(0)}{d\ln l}=\sum_z 2C[V_{1,-1}(z)]^2\left(K(0) - K(z)\right) ,
\end{equation}
%\begin{equation}
%\label{RG2FM}
%\frac{dK^{-1}(z)}{d\ln l} =-\frac{4C}{N} [V_{1,-1}(z)]^2\sum_{q_z} \tilde{K}(q_z)\sin^2\left(\frac{q_zz}{2}\right),
%\end{equation}
%where $z\neq 0$ and
%\begin{equation}
%\label{RG0M}
%\frac{dK^{-1}(0)}{d\ln l}=\frac{4C}{N} \sum_{z,q_z} [V_{1;-1}(z)]^2 \tilde{K}(q_z)\sin^2\left(\frac{q_zz}{2}\right).
%\end{equation}
and they should be considered self-consistently together with Eqs.(\ref{RG111M}, \ref{FurM}, \ref{Fu2rM}).
An elementary inspection of Eqs.(\ref{RG111M}, \ref{FurM}, \ref{Fu2rM}, \ref{RG2FM}, \ref{RG0M}) shows that $\tilde{K}(q_z=0)$  is not affected by the RG because $\sum_z \frac{dK^{-1}(z)}{d\ln l}=0$. This implies that the field $\Psi_{\rm CSF}=\psi_1 \psi_2...\psi_N$ always remains condensed. Furthermore, as long as the initial flow of $V_{1;-1}(z)$ (described by Eq.(\ref{RG22M})) drives the amplitudes away to $\infty$, the fixed point for the Luttinger matrix is given by $K(z)=K(0)$, that is, by $\tilde{K}(q_z\neq 0)=0$.

\smallskip
\subsection{RG for supercounterfluids} 
If there is a pair of tubes $z,z'$ ($z\neq z'$) with filling factors $\nu_z\neq 1/2$ and $\nu_{z'}=1-\nu_z$, the BS harmonic $V_{1;1}(z,z')\cos(2 \theta_{z} + 2 \theta_{z'}) $ can become relevant, while $V_{1;-1}(z,z')$ is irrelevant due to the mismatch of the filling factors. As a consequence, the gapless superflow is possible only in the counter-flow channel. In other words, it is the difference between the two phases which remains gapless.

The RG equations for the counterflow can be written for each pair $z,z'$ of tubes with the complementary filling factors by simply changing the sign in front of the $K_{zz'}$ term in the corresponding equations, Eq.~\eqref{RG11}, derived above for the complex superfluids (see details in the Appendix \ref{A}).
Specifically, we find
%\begin{widetext}
\begin{equation}
\label{RG11C}
\frac{dV_{1;1}(z,z')}{d\ln l }= \left[2- K_{zz} -K_{z'z'} - 2K_{zz'} \right] V_{1;1}(z,z')\;.
\end{equation}
%\end{widetext} 
Here, in full analogy with the composite superfluids, the $V_{1;1}$ channel can become gapped in the limit $V_d \to 0$. 

For the case of more than two tubes in the counterflow regime, the dipolar interaction can induce an additional gap in the $\sim V_{1;-1}$ channel in tubes with identical filling factors. However, a simple count of the remaining gapless phases shows that the gap in the $V_{1;-1}$ channel does not change their number. Indeed, let's consider two sets of tubes, $M_1>1$ and $M_2>1$, so that in the first one the filling factor in each tube is $\nu \neq 1/2$ and in the second one it is $1-\nu$. Then, there will be gaps in the channels $V_{1;-1}(z_1,z'_1)$ for each pair $z_1,z'_1$ from the first set of $M_1$ tubes and in $V_{1;-1}(z_2,z'_2)$ for each pair
$z_2,z'_2$ from the other set. As a result, there are two total phases from each group left gapless. Then, the channels   $V_{1;1}(z_1,z_2)$ also become gapped due to the counter-flow BS. This leaves just one phase gapless.  
The described situation  has a very simple interpretation: the gaps in tubes with equal filling factors imply formation of a {\it pair} of composite superfluids---one per each group of tubes and these composite superfluids further bind in the counterflow regime, as discussed in the section \ref{SCF}.

Similarly to the composite superfluids, Eq.~\eqref{RG2M}, the Luttinger matrix satisfies (see Appendix \ref{A})

\begin{equation}
\label{RG2C}
\frac{d(\hat{K}^{-1})_{zz'}}{d\ln l} =C [V_{1;1}(z,z')]^2(K_{zz} + K_{z'z'} + 2K_{zz'}),
\end{equation}
where $z\neq z'$, and
\begin{widetext}
\be
\label{RG2Cdiag}
\frac{d(\hat{K}^{-1})_{zz}}{d\ln l} =C \sum_{z'\neq z} \left\{ [V_{1;1}(z,z')]^2(K_{zz} + K_{z'z'} + 2K_{zz'})
+ [V_{1;-1}(z,z')]^2(K_{zz} + K_{z'z'} - 2K_{zz'})\right\}.
\ee
\end{widetext}
The first sum here is the contribution from the pairs of tubes with complementary filling factors, and the second one is due to the tubes with same filling factors.
 
Finally, we write the above equations for the case of translational symmetry along z-axis. This can be realized when, for example, tubes with even z-coordinates  ($z=0,2,4,...$) have filling factor $\nu$ and tubes with odd z ( $z=1,3,5,...$) have filling factor $1-\nu$ (see Fig.~\ref{fig:SCF}).
Then, similarly to the composite superfluid case 
\begin{equation}
\label{RG4}
\frac{dV_{1;1}(z)}{d\ln l}=2 \left[1- K(0) - K(z)\right] V_{1;1}(z),
\end{equation}
%\begin{equation}
%\label{RG4}
%\frac{dV_{1;1}(z)}{d\ln l}= \left[2- \frac{4}{N} \sum_{q_z} \tilde{K}(q_z) \cos^2\left(\frac{q_zz}{2}\right)\right] V_{1;1}(z),
%\end{equation}
where the distance $z=z -z'=1,3,5,...$ corresponds to pairs of tubes with the complementary filling factors.
For the distances $z=2,4,6,..$, that is, for layers with same filling factors, Eq.~\eqref{RG111M} has to be used.
Similarly the flow of the matrix of stiffnesses at odd distances $z$ is given by 
\begin{equation}
\label{RG3}
\frac{dK^{-1}(z)}{d\ln l} =2C [V_{1;1}(z)]^2(K(0) + K(z)),
\end{equation}
%\begin{equation}
%\label{RG3}
%\frac{dK^{-1}(z)}{d\ln l} =\frac{4C}{N} [V_{1;1}(z)]^2\sum_{q_z} \tilde{K}(q_z)\cos^2\left(\frac{q_zz}{2}\right),
%\end{equation}
while even, non-zero distances $z$ are described by Eq.~\eqref{RG2FM}, and the
diagonal term has contribution from all the pairs of tubes 
\begin{widetext}
\begin{equation}
\label{RG00}
\frac{dK^{-1}(0)}{d\ln l} =2C \sum_{z=1,3,5,...} [V_{1;1}(z)]^2(K(0)+K(z)) +2C\sum_{z=2,4,6,...} [V_{1;-1}(z)]^2 (K(0) - K(z)).
\end{equation}
%\begin{equation}
%\label{RG00}
%\frac{dK^{-1}(0)}{d\ln l} =\frac{4C}{N}\sum_{q_z}\tilde{K}(q_z)\left[ \sum_{z=1,3,5,...} [V_{1;1}(z)]^2 \cos^2\left(\frac{q_zz}{2}\right) + \sum_{z=2,4,6,...} [V_{1;-1}(z)]^2 \sin^2\left(\frac{q_zz}{2}\right)\right].
%\end{equation}
\end{widetext}

It is instructive to ignore Eq.~\eqref{RG3} and \eqref{RG00}, and substitute the initial value \eqref{K} into Eq.~\eqref{RG4} in the limit $V_d \to 0$. For $z\neq 0$ this gives
\begin{equation}
\label{RG44}
\frac{dV_{1;1}(z)}{d\ln l}\approx  \pi (\bar{V}(0)+\bar{V}(z)) V_{1;1}(z),
\end{equation}
where we have only used the first order term in $V_d \to 0$ while expanding \eqref{K}.
Thus, for purely repulsive interaction, Eq.(\ref{eq:Vdrep}), the harmonic $V_{1;1}(z)$ is relevant for arbitrary small $V_d$  in a direct analogy with the CSF case. Furthermore, it is interesting to note that inter-layer attraction $\bar{V}(z) <0$, Eq.(\ref{Vdip}), 
also induces the composite super-counter-fluid as long as the intra-layer repulsion is strong enough, that is,  $\bar{V}(0) > |\bar{V}(z) | $. 

The analysis of the above equations shows that for even number of layers, the fixed point corresponds to $ K(z)=K(0)$ for $z=2,4,6,...$ and $K(z)=-K(0)$ for $z=1,3,5,...$. Thus, while $\tilde{K}(q_z=0)=0$, the harmonic at $q_z=\pi$ remains condensed (because $\tilde{K}(q_z=\pi)\neq 0$). This, as discussed earlier, corresponds to the supercounterfluidity in the nearest neighbor layers, with the condensed field $\Psi_{\rm SCF}=\psi_1\psi^\dagger_2\psi_3\psi^\dagger_4...$ . 

\subsection{RG  for $\nu =1/2$ insulators} \label{section:insulators}
In the absence of inter-tube interactions, hard core bosons can form a checkerboard (CB) insulator at filling factor $\nu=1/2$ only if the repulsive interaction is strong enough, so that the Luttinger parameter $K$ is reduced from $K=1$ to $K=1/2$ (see Ref.~\cite{Giamarchi}). This can also be seen from Eq.~\eqref{RG11C} written for $z=z'$, that is, for the intra-tube harmonic $\cos(4 \theta_z + 4\pi \nu_z x)$. In this case, Eq.~\eqref{RG11C} becomes $dV_{1;1}(0)/d\ln l= (2- 4 K_{zz})V_{1;1}(0)$. In the absence of inter-tube interaction the Luttinger matrix becomes diagonal $(\hat{K})_{zz'}=K\delta_{zz'}$, so that one can write $dV_{1;1}(0)/d\ln l= (2- 4 K)V_{1;1}(0)$, implying that the critical value $K=K_c=1/2$. 

The situation changes dramatically in the presence of inter-tube interaction.
At filling $\nu =1/2$ harmonics $V_{1;1}(z), V_{1;-1}(z)$ with $z\neq 0$ can become relevant for $K=1$, as Eqs.(\ref{RG22M}, \ref{RG44}) indicate. This happens regardless of the sign of the inter-tube interaction even in the limit $V_d \to 0$ as long as $\hat{V}(0)>|\hat{V}(z)|$. Accordingly, all pairs of phases $\theta_z \pm \theta_{z'}$, with $z\neq z'$  become gapped, which implies that all the individual phases $\theta_z$ are gapped.

 It is possible to make a much stronger statement: for $N>2$, the $\nu=1/2$ insulating state occurs even in the absence of intra-tube interaction, i.e. $\bar{V}(0)=0$, and for purely attractive inter-tube interaction $\bar{V}(z) \to 0$. In order to see this, we analyze the RG Eq. (\ref{RG111M}) which, as the initial flow (\ref{RG22M}) indicates,
implies relevance of all $V_{1;-1}(z)$ for $z\neq 0$. Accordingly, as Eqs. (\ref{RG2FM}, \ref{RG0M}) show, the matrix $\hat{K}$ flows toward
 $\tilde{K}(q_z)=0$ (in the limit $\ln l \to \infty$)  for all $q_z$ except $q_z=0$. 
This means Eq.(\ref{RG4}) can be approximately rewritten as
\begin{equation}
\frac{dV_{1;1}(z)}{d\ln l} \approx [2- \frac{4}{N} \tilde{K}(0) ] V_{1;1}(z)
\label{RG4lim}
\end{equation}
in the limit $\ln l \to \infty$. Keeping in mind that at small $V_d \to 0$ the initial value $  \tilde{K}(0) \approx 1$, this equation shows that, even  if the renormalization of $\tilde{K}(0)$ is ignored, $ V_{1;1}(z)$ flows to $\infty$ as $ \sim  l^{2- 4/N}$ as long as $N>2$ (which means that the harmonic $q_z=0$ is also gapped and $ \tilde{K}(0)$ must actually flow to $0$).  %Further analysis of Eqs. (\ref{RG3}, \ref{RG00}) does show that this is indeed the case. 
As it will be seen below, this conclusion is also consistent with the simulations.

The $N=2$ case is special because, in the one-loop approximation, the RG equations for $V_{1;-1}$ and $\tilde{K}(\pi)$ are independent from the equations for $V_{1;1}$ and $\tilde{K}(0)$. Accordingly, the equation for $\tilde{K}(0)$ predicts that it must flow to a stable fixed point $\tilde{K}(0) > 1$ as long as $\bar{V}(0) < |\bar{V}(z)|$ and $\bar{V}(z)<0$ . %However, our simulations of the $N=2$ tube system for $\bar{V}(0)=0$ show that  $\tilde{K}(0)$ flows toward a superfluid fixed %point $\tilde{K}(0) \approx 0.5$. 
This issue will be discussed in greater detail elsewhere. Below we will explicitly demonstrate numerically  the thresholdless nature of the composite superfluid in the simplest case $N=2$.

%As already mentioned, in contrast to the filling factor $\nu =1/2$, forming insulating states at other special filling factors $1/3, 1/5, 2/5,$ etc. requires that interactions exceed certain thresholds so that initial values of the Luttinger parameter $K$ become significantly below unity. Particular dependencies of $K$ on $V_d$ are strongly non-perturbative and can only be established numerically. Similarly to these strongly interacting cases, the phases proposed in Ref.~\cite{Demler_2011} require {\it ab initio} simulations. Indeed, the BS harmonic responsible for binding one boson from one layer with $m $ bosons from another one is $V_{1,-m}(z,z')\cos[2 \theta_z - 2m \theta_{z'} + 2\pi (\nu_z - m \nu_{z'})x]$. It can be relevant, if, first, $\nu_z=m\nu_{z'}$ and, second, the Luttinger parameter becomes smaller than $K_c= 2/(1+m^2)$, as a simple analysis of the RG equations indicates. Thus, in order to realize cases with $m>1$ \cite{Demler_2011}, the inter-tube interaction should exceed some threshold to be found in the {\it ab initio} simulations.    

\section{Quantum Monte Carlo {\it ab-initio} algorithm and some results} \label{section:QMC}
The standard Worm Algorithm (WA) \cite{WA} is based on the evaluation of one-particle correlators $D_1$ in imaginary time and the possibility to switch effectively to the functional space of the partition function of the closed world-lines of particles. If $M=2,3,...$, particles (or holes) form bound complexes,  the efficient simulations can be achieved only through evaluation of the $M$-particle correlators. In this case $M=2$, such an algorithm has been developed in Refs.\cite{two_worms}. The situation becomes more complicated for $M>2$, when no effective switching to the partition function space can, in general, be achieved. This problem has been resolved in Ref.~\cite{Capogrosso} in the case of no inter-layer tunneling and in Ref.~\cite{Kuklov} in a more general setting. Here the algorithm \cite{Capogrosso} (designed to work in a discrete space-time) is extended to the quantum case, that is, to continuous time.    

While avoiding technical details, here we give a general overview of the quantities measured during the simulations. The most general correlator which can effectively describe a phase of $M$ bound complexes is the $M$-particle correlator ---a function of 6$M$ variables
\begin{equation}
D_M(r_1,..,r_M; r'_1,..,r'_N)=\langle A^\dagger(r_1,..,r_M) A(r'_1,..,r'_M)\rangle \; , 
\label{corr}
\end{equation}
where $\langle...\rangle$ stands for quantum-statistical averaging with the weight $\exp(-\beta H)$ determined by the Hamiltonian $H$ (\ref{eq:H}) and $A(r_1,..,r_M) = a(r_1)a(r_2) ...a(r_M)$, with $a(r_i)$ being bosonic annihilation operator in the space-time point $r_i=(x_i,z_i,\tau_i), $ with $i=1,2,...,M$. 
 The imaginary time dependence is given by the interaction representation defined for an operator $f$ as  $f(\tau) =e^{\tau H_0}f(0) e^{-\tau H_0}$, where $f(0)$ is the operator in the Schr\"odinger representation and $H_0$ is the part of the Hamiltonian which is diagonal in the Fock basis, that is, the interaction part of $H$. 

Evaluation of $D_M$ is based on the random walks of the $2M$ open {\it ends} of the world-lines, {\it worms} \cite{WA},  controlled by the famous Metropolis prescription. The identification of the phases, then, stems from the statistics of the relative distances between the worms, as  described in Ref.~\cite{Capogrosso}.    For example, in the CSF phase of complexes each composed of $M$ particles, all the correlators $D_{M'}$ (\ref{corr}) with $M'<M$ exhibit exponential decay with respect to all the pairs of space-time distances $r_i - r_j$,
$r'_i-r'_j$ and $r_i-r'_j$ where $i,j =1,2,..,M'$. This behavior is the key signature of  insulators with no off-diagonal long range (or algebraic) order.  A completely different behavior is demonstrated by the correlator $D_M$. On the one hand, if all the {\it ends} from one set, e.g., $r_1,r_2,...,r_M$ are kept inside a small region, the {\it ends} from the other set will automatically stay close together within some finite radius $\xi_0$ determining a typical extension of the constituents forming one complex, that is, $\langle |r'_i - r'_j| \rangle \leq \xi_0$. On the other hand, the dependence of $D_M$ on the relative space-time distance between the "centers of mass"
$|R_{cm}-R'_{cm}|$ (defined as $R_{cm} = [r_1 + ... + r_M]/M$ and $R'_{cm} = [r'_1 + ... + r'_M]/M$)
features the off-diagonal long range (or algebraic) order. The transition from CSF to the standard superfluid (where $D_1$ is long ranged) is marked by the divergence of $\xi_0$.

Keeping in mind the specificity of the present system, we evaluated the correlator $D_N$ and kept only one pair of the variables $r_i ,r'_i$, one from the first set and one from the other, in each tube (there is no inter-tube tunneling so that each  worm stays in its tube).   In order to realize the "confinement" of the first set of variables $r_1,...r_N$,  we have introduced an artificial configuration weight $W\sim \exp [-\sum_{m,n}^N(|x_m-x_n|/\xi_x +|\tau_m-\tau_n|/\xi_\tau)]$, where $\xi_x $, $\xi_\tau$ are  microscopic parameters chosen so that as to maximize the algorithm efficiency. Accordingly, the expectation values $\langle ... \rangle$ are evaluated with respect to the weight $W \exp(-\beta H)$.

As demonstrated in Refs.\cite{Capogrosso,Kuklov}, the described approach turned out to be very effective in idenifying various phases as well as the universalities of the transitions. It can be easily adjusted to various systems. For example, if considering the bilayer system proposed in Ref.~\cite{Burovski} the correlator $D_{p+q}$ should be used with $p$ pairs of the ends kept in one tube and $q$ pairs --- in the other.

Here we present  results of {\em ab-initio} Quantum Monte Carlo (QMC) simulations based on the path integral with $H$ from Eq.(\ref{eq:H}) and  focusing on demonstrating explicitly the absence of the threshold for the formation of the CSF state. Unless otherwise noted the simulations were performed in the case of nearest neighbor inter-layer attraction, and in the absence of intra-layer interactions. Specifically, we considered the case $N=2$ and compared the result for the renormalized Luttinger parameter determined numerically (through the representations (\ref{R},\ref{C}, \ref{Lut})) with the prediction of RG. We have also performed simulations of the $N=3$ case within the approach described above and have demonstrated: 1.The formation of the CSF phase; 2. The existence of the insulating CB state of the chains at the filling $\nu=1/2$ and provided data consistent with the absence of the threshold for its formation.

\subsection{QMC study of the bilayer system, $N=2$ case}
Absence of the threshold for the phases discussed above implies that, in order to realize them, there is no need to pursue strong dipole-dipole interactions. Instead, the size of the system should be made large enough (and temperature low enough) so that the effects of small gaps are seen. Here we will address the issue of no threshold in detail by {\it ab-initio} simulations of the bilayer system. The goal of this study is to demonstrate this property explicitly.

We consider two identical layers located at $z=0,1$ with $\nu_0=\nu_1=\nu$. Then, the Luttinger matrix consists of just two elements $(\hat{K}^{-1})_{00}= (\hat{K}^{-1})_{11}$ and $(\hat{K}^{-1})_{01}$. Accordingly, the Fourier representation along the z-axis has just two harmonics with $q_z=0, \pi$, so that Eq.~\eqref{Fu2rM}) yields $1/\tilde{K}(0)= (\hat{K}^{-1})_{00} + (\hat{K}^{-1})_{01}$ and $1/\tilde{K}(\pi)= (\hat{K}^{-1})_{00} - (\hat{K}^{-1})_{01}$. As presented in Eqs.~\eqref{FurR}, \eqref{FurC}, \eqref{Lut}
\be
\label{pi}
\tilde{ K}(\pi)=\frac{\pi}{2}\sqrt{\langle (W_x(0) - W_x(1))^2 \rangle \langle (W_\tau (0) - W_\tau (1))^2\rangle},
\ee
\be
\label{0}
\tilde{K}(0)=\frac{\pi}{2}\sqrt{\langle (W_x(0) + W_x(1))^2 \rangle \langle (W_\tau (0) + W_\tau (1))^2\rangle},
\ee
in terms of space-time windings $W_x(0)$, $W_x(1)$, $W_\tau (0)$, $W_\tau (1)$.  

We have determined  $\tilde{K}(\pi)$ by QMC for various interactions and system sizes, where the RG scale $ l$ was identified with the system size $L$, provided the inverse temperature $\beta=1/T \propto L$ in the atomic units. Practically,  we have kept $L \propto \beta$ so that $\langle (W_x(0) - W_x(1))^2 \rangle =\langle (W_\tau (0) - W_\tau (1))^2\rangle$, in order to ensure space-time symmetry, that is, that the system is in its ground state.
\begin{figure}[h]
\begin{center}
\includegraphics[width=0.5\textwidth]{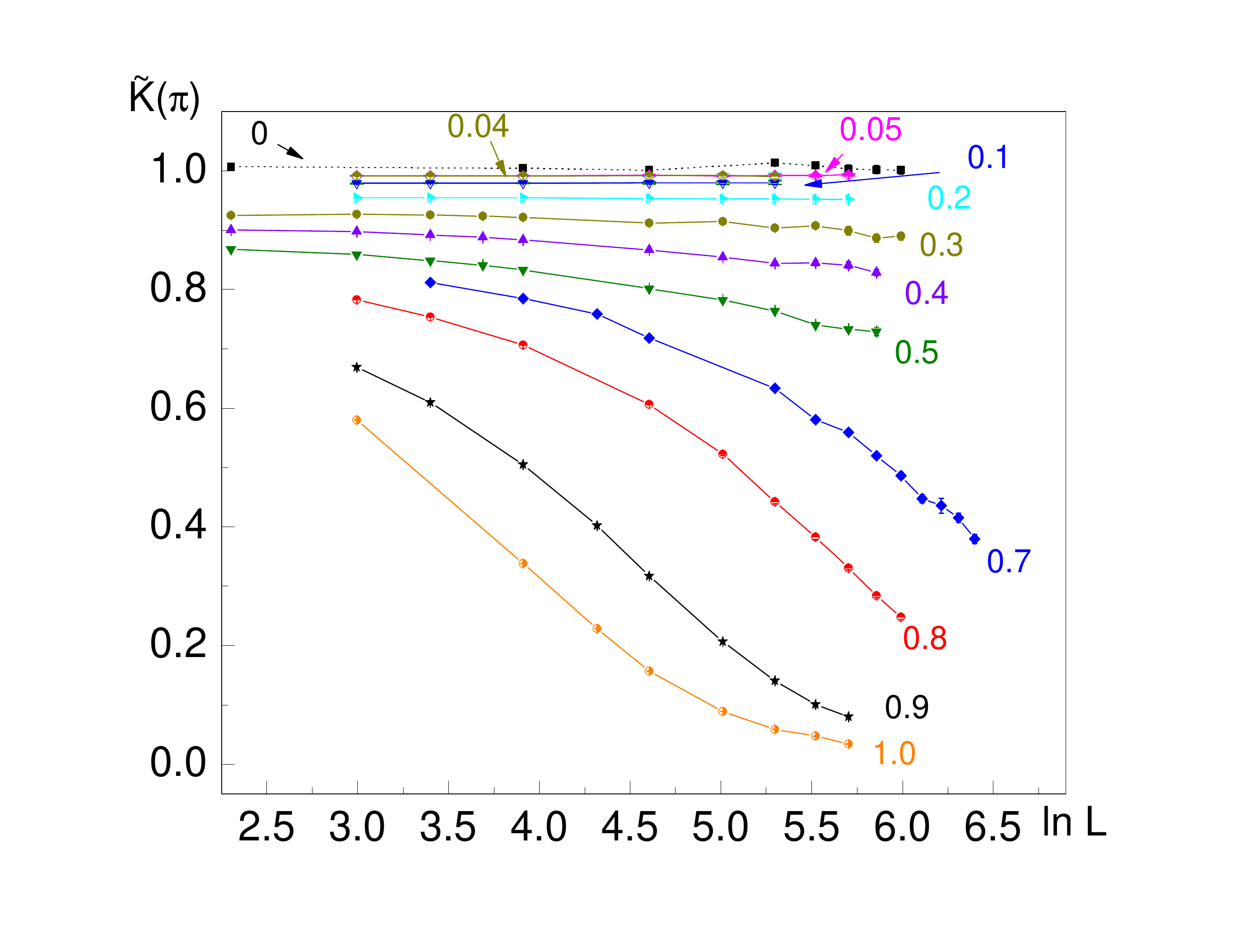}
\caption{(Color online) Numerical data for $\tilde{K}(\pi)$ as a function of system size $L$ for different values of the inter-layer interaction $V_d/J$, and in the absence of intra-layer repulsion.  }
\label{fig:K-N2}
\end{center}
\end{figure}
\begin{figure}[h]
\begin{center}
\includegraphics[width=0.5\textwidth]{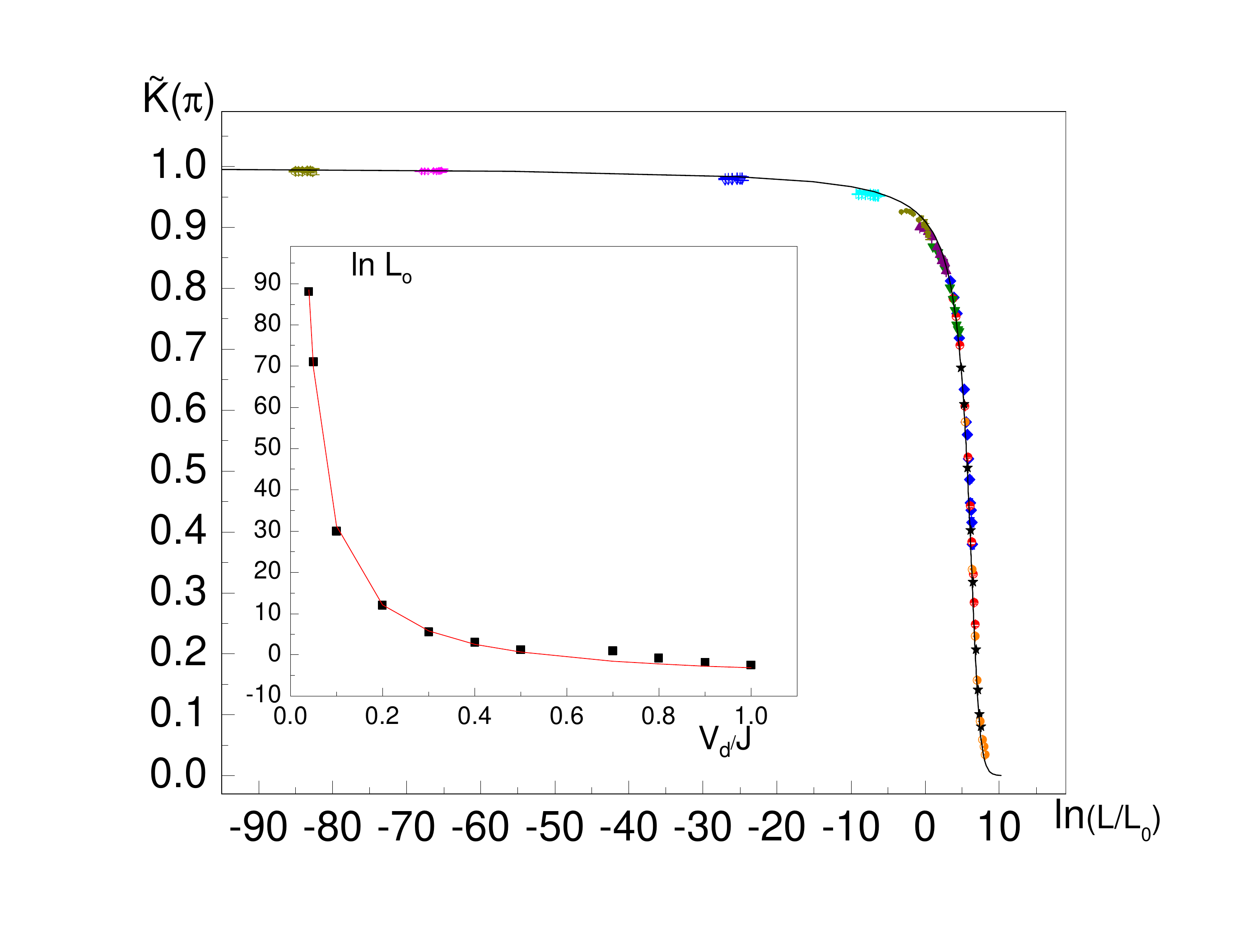}
\caption{(Color online) The data from Fig.~\ref{fig:K-N2} is shown vs $\ln(L/L_0(V_d))$ with $L_0(V_d)$ chosen in such a way as to achieve the collapse on a single curve. The size of the symbols is determined by the statistical error bars. The solid line is the RG solution for the separatrix with the critical value of the Luttinger parameter being $K_c=1$, Eq.~(\ref{sepM}). Inset: plot of the rescaling parameter $\ln L_0$ vs  $V_d/J$. Solid line is the fit by $\ln L_0 = a/(V_d/J)  - b$, with  $ a=3.82, b = 6.96$.  }
\label{fig:master}
\end{center}
\end{figure}
Our purpose is comparing the numerical dependancies of $  \tilde{ K}(\pi)$ vs $L$ for various interaction strengths $V_d$ with the RG flows. The raw data for  $\tilde{ K}(\pi)$ is presented in Fig.~\ref{fig:K-N2}.
As it turned out, within the statistical errors of the simulations, the curves of $\tilde{ K}(\pi)$ vs $L$ for various $0< V_d/J < 1$ have been found to  belong to one master curve ---the separatrix of the RG equations \eqref{eq:sol}-\eqref{sep}  (discussed in the Appendix B), which can be represented as
\be
\label{sepM}
\ln \xi_s -\frac{1}{\xi_s}=2\ln\left(\frac{L}{L_0}\right),\quad \xi_s=\frac{1}{\tilde{K}(\pi)} -1.
\ee
where $L_0(V_d/J)$ is a rescaling parameter which can be interpreted as the length $\xi_0$ --- the size of a bound dimer.
This dependence has been found from  rescaling $\ln L \to \ln L -\ln[ L_0(V_d)]$ for each value of the interaction. The result of this procedure is presented in Fig.~\ref{fig:master}.  As can be seen from the inset, $L_0$ as a function of the inter-tube interaction diverges as
\be
\label{corM}
L_0 \sim \exp\left(\frac{aJ}{|V_d|}\right),\quad V_d \to 0
\ee
where $a$ is a constant ( $a=3.82$). Such a divergence proves that the critical value for the formation of the dimer superfluid is $V_d=0$. Thus,  the accurate matching of the numerical data by the RG solution (\ref{sepM}) over almost 50 orders of magnitude of the (effective) distances as well as the dependence (\ref{corM}) indicate that paired superfluid is formed for infinitesimally small inter-layer interaction strength.
 Such an approach -- matching numerical solution by the RG flow for finding critical point of Berezinskii- Kosterlitz-Thouless  transition \cite{BKT} -- has been pioneered in Ref.\cite{Borya_Kolya}.

\subsection{QMC results for $N=3$ tubes}

Below we present QMC results obtained by the multi-worm algorithm for the case of $N=3$ tubes.
 As it has already been mentioned, achieving efficient numerical convergence  by the approach \cite{two_worms} in the cases $N>2$ is not possible. Instead, the simulations should focus on evaluating  the $N-$body correlator $D_N$, Eq.(\ref{corr}). Then, the determination of the phases can be based on observing spatial dependencies of the corresponding correlators.

 We introduce two quantities $f_1(x'_1-x'_2)$ and $f_2(x_1 -x'_1)$ which can be viewed as spatial projections of the full correlator $D_3$ where the pair $x'_1,x'_2$ in $f_1$ belongs to the "primed" coordinates in the definition (\ref{corr}), and $x_1,x'_1$ in $f_2$ are from the "unprimed" and the "primed" sets, respectively. Specifically,
$
f_1(x'_1,x'_2)\propto\int d\tau'_1 d\tau'_2dr_1 dr_2 dr_3 dr'_3 D_3 W
$
 and $f_2(x_1,x'_1)\propto\int d\tau_1 d\tau'_1 dr_2 dr_3 dr'_2 dr'_3D_3 W$, with the artificial weight $W$ discussed at the beginning of the section  \ref{section:QMC}. 

\begin{figure}[h]
\begin{center}
\includegraphics[width=0.5\textwidth]{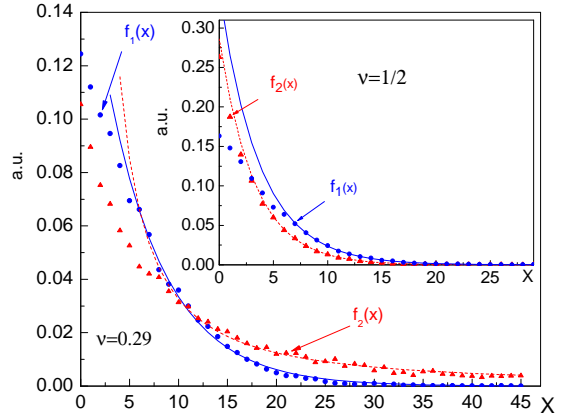}
\caption{The CSF and CB phases in the $N=3$ tubes: QMC data (points) and their fits  (lines) in the presence of dipolar interaction. The filling factors are shown.
Main pannel: while $f_1(x)$ exhibits the exponential decay $\sim \exp(-0.169|x|)$, $f_2(x)$ shows algebraic order $\propto |x|^{- 1.39}$. Inset: in the CB phase both functions are exponentially decaying, $f_1\sim \exp(-0.269|x|), \,\, f_2 \sim \exp(-0.310 |x|)$.}
\label{fig:D3}
\end{center}
\end{figure}

Given the definition,  $f_1(x)$ must exhibit exponential decay in the $N=3$ CSF phase as well as in the insulating phases.
The function $f_2$, while demonstrating the exponential decay in the insulator, should show algebraic behavior in the CSF phase.
These features are clearly seen in Fig.~\ref{fig:D3} for $N=3$ identical tubes for two filling factors and in the presence of the full dipolar interaction at $V_d/J=0.75$.
While the main plot clearly shows the CSF ($\nu =0.29$), the inset represents the CB insulator ($\nu=1/2$).  
\begin{figure}[h]
\begin{center}
\includegraphics[width=0.5\textwidth]{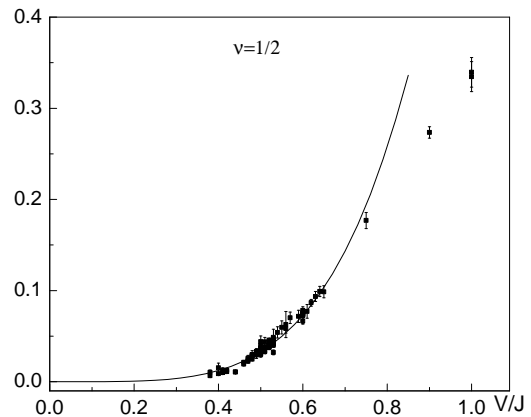}
\caption{The CB contrast for $N=3$ plotted for all sizes $L=100-600$, in the absence of intra-layer repulsion and all inter-layer interaction strengths $V_d/J$ studied. Solid line is the fit by the power-law dependence $\propto V^{4.39}_d$.}
\label{fig:str}
\end{center}
\end{figure}

The CB phase is characterized by finite amplitude of the modulation of the density $\rho (x) = \sum_{z} n_z(x)$ at the wavevector $\pi$. 
The RG analysis conducted  in Sec.~\ref{section:insulators} indicates that such modulation can occur even in the absence 
of the intra-layer repulsion due to arbitrary small inter-layer attraction $V_d$. In other words, the repulsive interaction causing the CB order is to be induced dynamically even if it is not present microscopiocally.  This analysis, however, does not predict strength of such interaction. In our simulations without any intra-layer repulsion we were able to resolve the CB modulation only for $V_d/J \geq 0.38$. Furthermore,
the structure factor correlator $S(x) =(-1)^x \langle \rho(x) \rho(0) \rangle$ showed a very weak dependence on the system size $L$ for the whole range of measurements $0.38 \leq V_d/J \leq 1$. In these circumstances conducting the comparison with the RG flow like it was done for the case of the paired superfluid does not appear to be feasible. In other words, a weak dependence of the induced repulsion on $V_d$ does not allow approaching the critical region at small $V_d$. Thus, the observed CB order corresponds to the values of the renormalized Luttinger parameter which are already so small that the structure factor $\sim C_2 |x|^{-2K}$ becomes essentially independent of $x$, with the factor $C_2$ being a non-universal coefficient (cf. with the spin$=1/2$-chain magnetization modulation in Sec. 6 of Ref.\cite{Giamarchi}). However, despite such limitations, there is a feature which is consistent with the thresholdess nature of the CB. Fig.~\ref{fig:str}   shows  onsite  CB contrast measured for all the system sizes and the inter-layer strength studied. The data can be fit by the power law dependence on $V_d$, $C_2 \propto V_d^b, \, b\approx 4.39$, which is consistent with no threshold in $V_d$.

\section{Conclusions and outlook}
Superfluids of hard-core bosons in the multi-tube geometry turn out to be unstable toward forming composite superfluids, supercounterfluids and CB insulators for arbitrary small inter-tube interaction. This conclusion is supported by the bosonization and numerics based on the newly developed {\it ab initio} Monte Carlo alghorithm. Thus, while realizing experimentally such phases, the smallness of the dipolar interaction can be to some extent compensated by enlarging  system size.

In the context of the emergence of  parafermions \cite{Kuklov} in the multi-tube geometry with finite inter-tube tunneling, we find important conducting {\it ab initio} simulations of such systems in order to establish the requirements for their experimental realization. In particular, such simulations are needed to infer how the threshold for the transitions depends on the dipolar strength and the tunneling amplitude.
       
Another interesting system proposed in Refs.\cite{Burovski, Demler_2011} also requires {\it ab initio} simulations for establishing  practical ranges of the interaction and lattice parameters. As bosonization argument indicates, forming a superfluid consisting of complexes of $p \geq 1$ hard core bosons from one tube with $q>1 $ such bosons from the other requires exceeding some finite threshold in $V_d$. Indeed, the critical value of the Luttinger parameter needed to make the BS harmonic $V_{p,-q}$ (\ref{BS}) relevant is $K_c= 2/(p^2 + q^2)$, which is significantly smaller than $K=1$ for the hard core bosons even for the lowest non-trivial combination $p=1,q=2$.  Increasing the intra-layer interaction reduces the $K$ value, so that, potentially, it may be possible to realize the phases \cite{Burovski, Demler_2011}. This, however, needs to be checked by the QMC.

{\bf Acknowledgement} --
this work was supported by the NSF through a grant to ITAMP at the Harvard-Smithsonian Center for Astrophysics,  grant PHY1314469, and by the grant from CUNY HPCC  under NSF Grants CNS-0855217, CNS-0958379 and ACI-1126113.

\begin{appendix}
\section{RG equations}\label{A}
Here we outline derivation of the RG equations for general (weak) interactions.
The procedure is a straightforward extension of the standard one (see, e.g., Refs.~\cite{Giamarchi,Caza}).   

For small interactions, the only relevant harmonics in the backscattering terms (\ref{BS}) can be those with the lowest integers $m_z, m_{z'}$, that is, $m_z=\pm 1$, $m_{z'}\pm 1$, with $z\neq z'$ . The standard renormalization procedure consists of integrating out small oscillations of the Haldane phases $\theta_z$ \cite{Haldane_1980} (from the partition function $Z =\int D\theta \exp(-S)$) within the spherical shell of $\vec{q}$ between some cutoff $\Lambda/(1+s)$ and $\Lambda$, and further rescaling $x \to (1+s) x$ and $\tau \to (1+s)\tau$ , with $s\to 0$. In the lowest order (one-loop approximation), this procedure implies independent renormalization of each harmonic. Specifically, for the case of $V_{m,-m}$ one finds
\begin{equation}
\label{RG1}
\frac{dV_{m;-m}(z,z')}{ds}= \left[2- \frac{2m^2}{s} \langle (\theta_z -\theta_{z'})^2 \rangle' \right] V_{m;-m}(z,z')
\end{equation}
where $\langle ... \rangle'$ implies Gaussian average with respect to the action (\ref{Rgauss}) , with the integration performed over the shell of the momenta defined above. In the $D=1+1$ dimensions, $\langle (\theta_z -\theta_{z'})^2 \rangle'$ exhibits log-divergence, that is, $\langle (\theta_z -\theta_{z'})^2 \rangle' \sim \ln(1+s) \sim s$ and it is independent of $\Lambda \to 0$. Then, the RG flow is controlled by $\ln \Lambda $ or, in a finite system of size $L$, by $\ln L$ so that $d.../ds =d .../d\ln L$. 

The mean $\langle ...\rangle'$ can be represented in terms of the elements of the Luttinger matrix $\hat{K}$ from Eq.(\ref{Rphi}) which is the inverse of the matrix $\hat{K}^{-1}$ from the dual form (\ref{Rgauss}). Thus, Eq.~(\ref{RG1}) becomes
\begin{widetext}
\begin{equation}
\label{RG11A}
\frac{dV_{m;-m}(z,z')}{ds}= \left[2- m^2 (K_{zz} + K_{z'z'} - 2K_{zz'}) \right] V_{m;-m}(z,z'),
\end{equation}
\end{widetext}
where $K_{zz'}$ are elements of the matrix $\hat{K}$.
In the case of translational invariance along the $z$-axis this equation can be explicitly written in the form (\ref{RG111M}).

%\begin{equation}
%\label{RG111}
%\frac{dV_{m;-m}(z)}{ds}= \left[2- \frac{4m^2}{N} \sum_{q_z} \tilde{K}(q_z) \sin^2\left(\frac{q_zz}{2}\right)\right] V_{m;-m}(z).
%\end{equation} 
%In the above expression we have used $V_{m;-m}(z,z')=V_{m;-m}(z-z')$ and $(\hat{K}^{-1})_{zz'}=\hat{K}^{-1}(z-z') $, with the inverse Fourier for the Luttinger matrix defined as:
%\begin{equation}
%\label{Fur}
%K_{zz'}=K(z-z')= \frac{1}{N} \sum_{q_z} \tilde{K}(q_z){\rm e}^{iq_z (z-z')}.
%\end{equation}
%Similarly, the inverse Fourier of the inverse Luttinger matrix is given by:
%\begin{equation}
%\label{Fu2r}
%(\hat{K}^{-1})_{zz'}=\hat{K}^{-1}(z-z')= \frac{1}{N} \sum_{q_z} (\tilde{K}(q_z))^{-1}{\rm e}^{iq_z (z-z')}.
%\end{equation}

If one ignores the renormalization of the Luttinger matrix, the value of $\tilde{K}(q_z)$ from Eq.~\eqref{K} can be used. For small $V_d$ one can expand Eq.~\eqref{K} in powers of $V_d$ and rewrite Eq.(\ref{RG11A}) as
\begin{widetext}
\begin{equation}
\label{RG22}
\frac{dV_{m;-m}(z)}{ds}= \left[2- m^2\left(2K - \pi K^2(\bar{V}(0)- \bar{V}(z))\right)\right] V_{m;-m}(z),
\end{equation}
\end{widetext}
where $\bar{V}(z)$ is given in Eq.~\eqref{Vdip}). In the limit $V_d \to 0$ and for $\vert m\vert=1$, the critical value of $K$ is $K_c=1$.
 We also note that higher harmonics $ V_{m,-m}(z), \, m>1,$ are irrelevant because the critical value for them in the limit $V_d\to 0$ is $K_m = 1/m^2 <1$.

The renormalization of the BS amplitudes, Eq.~\eqref{RG1}), is considered together with the renormalization of the inverse of the matrix $(\hat{K}^{-1})_{zz'}$ entering the quadratic form \eqref{Full}. In the one-loop approximation the main contribution is due to the same BS harmonic, $\cos(2(\theta_z \pm \theta_{z'}))$. It generates the term $\sim [\vec{\nabla}(\theta_z \pm \theta_{z'})]^2$ in the second order with respect to the harmonics $\theta'$ belonging to the RG shell, where the signs $\pm$ are correlated. Thus, the contributions to the diagonal elements $(\hat{K}^{-1})_{zz} $ and to the off-diagonal ones $(\hat{K}^{-1})_{zz'}$ where $z\neq z'$, should be considered independently.
Following the standard procedure (see in Refs.\cite{Giamarchi,Caza}),
the contribution to $(\hat{K}^{-1})_{zz'}$ from the BS amplitude $V_{1;\pm 1}(z,z')$, with $z\neq z'$, can be represented as 
\begin{equation}
\label{RG1K}
\frac{d(\hat{K}^{-1})_{zz'}}{ds}= \pm C V^2_{1;\pm 1}(z,z')  \frac{\langle (\theta_z \pm \theta_{z'})^2 \rangle'}{s} ,
\end{equation}
where the signs "$\pm$ " are correlated; $C$ is a non-universal constant determining type of the short distance cut-off (see in Ref.~\cite{Giamarchi}).

 The contributions to $ \frac{d(\hat{K}^{-1})_{zz}}{ds}$ come from all pairs. Specifically, 
\begin{widetext}
\begin{equation}
\frac{d(\hat{K}^{-1})_{zz}}{ds}= C \sum_{z'\neq z}\left[  V^2_{1;-1}(z,z')  \frac{\langle (\theta_z - \theta_{z'})^2 \rangle'}{s} 
+  V^2_{1;1}(z,z')  \frac{\langle (\theta_z + \theta_{z'})^2 \rangle'}{s}\right].
\label{KZZ}
\end{equation}
\end{widetext}
 Eqs.~(\ref{RG1K},\ref{KZZ}), where $\langle ... \rangle'$ implies averaging over the gaussian fluctuations within the momentum shell,  lead to Eqs.(\ref{RG2M},\ref{RG2dM},\ref{RG2C},\ref{RG2Cdiag}), where it is taken into account that $  \pm \langle \theta_z \theta_{z'} \rangle' \sim \pm s K_{zz'}$.

\section{RG solutions for $N=2$}

At $\nu \neq 1/2$ the only relevant harmonic is $V_{1;-1}(1)$. Thus the RG flow affects $\tilde{K}(\pi)$ and $V_{1;-1}$ only. The corresponding RG equations follow from Eqs.~\eqref{RG111M},\eqref{RG2FM}, \eqref{RG0M} as 
\begin{align}
\label{eq:RGg}
\nonumber \frac{d u}{d\ln l}&=2(1-g)u\\
\frac{dg^{-1}}{d\ln l}&=g u^2\;, 
\end{align}
where we used the notations $u=\sqrt{2C}V_{1;-1}(1)$, $g=\tilde{K}(\pi)$. These equations are the standard Kosterlitz-Thoulless \cite{BKT} RG equations (see in Refs.\cite{Giamarchi,Lubensky}). 

The flow $g(l)$ begins at small scales from the initial value
set by $g(0)=K/\sqrt{ 1+ \pi K \tilde{V}(\pi)}$, with $ \tilde{V}(\pi)= \bar{V}(0) - \bar{V}(1)$. Thus, $g(0) <1$ is below the critical value $K=1$
and the system should become gapped. 

The channel $V_{1;1}, \tilde{K}(0)$ is irrelevant as long as $\nu \neq 1/2$. At $\nu =1/2$, or in the case of the complementary filling factors $\nu_0 = \nu$ and $\nu_1 = 1-\nu$, the channel $(V_{1;1},\, \tilde{K}(0))$ must be considered as well.
The corresponding RG equations follow from Eqs.~\eqref{RG4} and Eqs.~\eqref{RG3}, \eqref{RG00} in the same form as Eqs.\eqref{eq:RGg} where
now $u=\sqrt{2C}V_{1;1}(1)$, $g=\tilde{K}(0)$, with the initial value set as $g(0) =K/\sqrt{ 1+ \pi K \tilde{V}(0)}$, with $ \tilde{V}(0)= \bar{V}(0) + \bar{V}(1)$. 
Thus, in the case $N=2$, the channels $V_{1;1}$ and $V_{1;-1}$  are decoupled from each other and are described by the same set of equations.

A general solution of the system \eqref{eq:RGg} can be expressed in terms of two constants of integration,  $\eta$, $l_0>0$, 
 determined by the initial values of $u$ and $g$, which in their turn are set by the microscopic model \eqref{eq:H}. If $\eta$ is real, the solution has a form
\begin{eqnarray}
\label{eq:sol}
u^2&=& 2[ \xi^2 -\eta^2],\, \xi = \frac{1}{g}-1, \, F_\eta=4\ln\left(\frac{l}{l_0}\right)\\
F_\eta&\equiv &\ln(\xi^2 (l) - \eta^2 ) + \frac{1}{|\eta| }\ln\left(\frac{\xi (l) - |\eta|}{\xi (l) + |\eta|}\right),
\nonumber
\end{eqnarray}
where $|\xi| > |\eta|$ and $\xi > -1$. If $\eta= {\rm i} |\eta|$, the solution becomes 
\begin{eqnarray}
\label{eq:sol2}
u^2&=& 2[ \xi^2 +|\eta|^2],\,\xi = \frac{1}{g}-1, \, F_\eta=4\ln\left(\frac{l}{l_0}\right), \\
F_\eta &\equiv&\ln(\xi^2 (l) + |\eta|^2 ) - \frac{2}{|\eta| }\tan^{-1}\left(\frac{|\eta|}{\xi}\right),
\nonumber
\end{eqnarray}
where $\xi > -1$.

 The constants $\eta, l_0$ are determined by the dipolar interaction, $V_d$. If $V_d=0$, that is, the hard-core bosons are non-interacting (except for the hard-core constraint), the RG equations are trivially satisfied by $\xi =0, u=0$, which implies that  $\eta=0, l_0 =\infty$ for $V_d=0$. 
The critical solution ($\xi=0, u=0$) belongs to the separatrix, $\eta \to 0$, $\xi(l)= \xi_s(l)$, $u(l)=u_s(l)$: 
\be
\label{sep}
u_s=\sqrt{2}|\xi_s|, \quad \ln \xi_s -\frac{1}{\xi_s}=2\ln\left(\frac{l}{l_0}\right).
\ee
Algebraic order exists in the domain $ -1< \xi(0) <0,\, u(0) <u_s$, where $\xi$ flows to the stable fixed point $\xi(\infty) = -|\eta|$ for real $\eta$ satisfying $0<|\eta| <1$, and $u(\infty)=0$. All other initial values correspond to the runaway flows $\xi(\infty)=\infty,\, u(\infty)=\infty$, that is, to the gapped state.

As explicitly shown in Eq.\eqref{K}, small inter-tube attractive interaction $V_d$ lowers $\tilde{K}(\pi)$ below $K=1$, that is, the initial value of $\xi$ is $\xi(0) \sim |V_d|$. It is also clear that the initial BS interaction $V_{1;-1}$ must also be $\sim |V_d|$ in this limit. Thus, $|\eta| \sim |V_d|$, as follows from Eqs.\eqref{eq:sol},\eqref{eq:sol2}. 

It is instructive to discuss the dependence $l_0$ vs $V_d$. As mentioned already, $l_0=\infty$ for $V_d=0$ and it must become finite as $V_d \neq 0$. Thus, $l_0$ has a meaning of the correlation length --- the size of a dimers forming paired superfluid. The type of the dependence can be established from, e.g., Eq.\eqref{eq:sol2}. Starting from $\xi(0) \sim |V_d|$ at $l \sim 1$, this equation becomes $ - |V_d|^{-1} \tan^{-1}(\kappa) \approx 4 \ln(1/l_0)$, where
$\kappa = |\eta|/\xi(0)$ is some number of the order of unity. Thus,
\be
\label{cor}
l_0 \sim \exp\left(\frac{\kappa'}{|V_d|}\right),
\ee   
where $\kappa' \sim 1$. As found in our simulations, Eq.~(\ref{corM}), this length, $L_0=l_0$, determines the properties of the paired superfluid. The dependence (\ref{cor}) should be, on one hand, contrasted with the temperature divergence $\sim \exp( ...1/\sqrt{T-T_c})$ of the correlation length in classical BKT transition on the approach to the critical temperature $T_c$ \cite{BKT}, and, on the other, it should be compared with the divergence of the two-body bound state size $\sim \exp(...1/ V_b)$ in 2D as the attractive potential $V_b \to 0$.

\end{appendix}

\end{document}